\documentclass[twocolumn,secnumarabic,amssymb, nobibnotes, aps, prd, cite
superscriptaddress,10pt]{revtex4}

\usepackage[utf8]{inputenc}
\usepackage{graphics,graphicx}
\usepackage{amsmath,amssymb,amsfonts,latexsym,cancel}
\usepackage[colorlinks]{hyperref} 
\usepackage{bm}

\begin{document}

\title{Relativistic polytropic spheres with electric charge: Compact stars,
compactness and mass bounds,  and quasiblack hole configurations}

\author{Jos\'e D. V. Arba\~nil}
\affiliation{Departamento de Ciencias, Universidad Privada del Norte, 
Avenida Alfredo Mendiola 6062 Urbanizaci\'on Los Olivos, Lima,  Peru.
Email: jose.arbanil@upn.pe
}
\author{Vilson T. Zanchin}

\affiliation{Centro de Ci\^encias Naturais e Humanas, Universidade Federal 
do ABC, Avenida dos Estados 5001, 09210-580 Santo Andr\'e, S\~ao Paulo, 
Brazil.
Email: zanchin@ufabc.edu.br  }

\begin{abstract} We study the static stellar equilibrium configurations of 
uncharged and charged spheres composed by a relativistic polytropic fluid, 
and compare with those of spheres composed by a non-relativistic polytropic 
fluid, the later case already being studied in a previous work
[J. D. Arba\~nil, P. S. Lemos, V. T. Zanchin,  Phys. Rev. D \textbf{88}, 
084023 (2013)]. 
For the two fluids under study, it is assumed an equation of state 
connecting the pressure $p$ and the energy density $\rho$. In 
the non-relativistic fluid case, the connection is through a non-relativistic 
polytropic equation of state, 
 $p=\omega\rho^{\gamma}$, with $\omega$ and $\gamma$ being respectively the 
polytropic constant and the polytropic exponent. In the relativistic 
fluid case, the connection is through a relativistic polytropic equation of 
state, $p=\omega\delta^{\gamma}$, with $\delta=\rho-p/(\gamma-1)$, and
$\delta$ being the rest mass density of the fluid. 
For the electric charge distribution, in both cases, we assume that the 
charge density $\rho_e$ is proportional to the energy density $\rho$, $\rho_e 
= \alpha\, \rho$, with $\alpha$ being a constant such that $0\leq 
|\alpha|\leq 1$. 
The study is developed by integrating numerically the hydrostatic 
equilibrium equation, i.e., the modified Tolman-Oppenheimer-Volkoff equation
for the charged case. Some properties of the charged 
spheres such as mass, total electric charge, radius, redshift, and the speed 
of sound are analyzed. The dependence of such properties with the polytropic 
exponent is also investigated. In addition, some limits that arise in
general relativity, such as the Chandrasekhar limit, the
Oppenheimer-Volkoff limit, the Buchdahl bound and the Buchdahl-Andr\'easson
bound, i.e., the Buchdahl bound for the electric case, are studied. As in a 
charged non-relativistic polytropic sphere,
the charged relativistic polytropic sphere with 
$\gamma\to\infty$ and $\alpha \to 1$
saturates the Buchdahl-Andr\'easson bound, thus indicating that it reaches
the quasiblack hole configuration. We show by means of numerical analysis that, 
as expected, 
the major differences between the two cases appear in the high energy density
region.

\end{abstract}
\maketitle

\section{Introduction}\label{sec-introd}

\subsection{Uncharged spheres: Equations of state and mass bounds}

In the study of stars, both in Newtonian gravitation and in General
Relativity, it is usual to model the matter inside the star by a perfect
fluid. Such a fluid is fully characterized by its energy density  $\rho$ and
pressure $p$, besides the speed of sound in it. In general, to close the
system of equations, and additional relation is needed. Usually, an
equation of state relating the pressure to the energy density of a fluid
in a very simple way is specified. Since Eddington \cite{eddington}, a
polytropic equation of state has been assumed to build
analytically simple star models. Such an equation relates the pressure and
energy density by a power law of the form
\begin{equation}
 p=\omega\rho^{\gamma}, \label{EoS1}
\end{equation}
where $\omega$ and $\gamma$ are respectively the polytropic constant and the 
polytropic exponent. Such a relation, which we call EoS $1$, is derived in 
Newtonian fluid mechanics, 
in which case $\rho$ is the mass density, but it is a good approximation for 
relativistic fluids as long as the energy density is sufficiently small (see, 
e.g., \cite{thorne_review}).

The equation of state \eqref{EoS1} has been used in several contexts. A 
fact of  interest here is that the first bound for the mass of a 
compact 
object was established, when studying white dwarfs, by using such a polytropic 
equation of state \cite{chandra1,chandra3}. In order to study the 
configurations of white dwarfs composed by a relativistically degenerate 
electron gas in a very simple manner, Chandrasekhar \cite{chandra1,chandra3} 
used EoS $1$ [see Eq. \eqref{EoS1}] with $\gamma=4/3$. Applying the laws of 
Newtonian gravitation, he found that the radius of the configuration decreases 
with growing of the energy density, and it shrinks to zero for a mass of 
approximately $1.44\,M_{\odot}$. This is the Chandrasekhar limit.

As in Newtonian gravitation, in the context of general relativity there are
also mass bounds for compact objects. Studies in this direction were
performed by Tolman \cite{tolman} and Oppenheimer and Volkoff
\cite{oppievolkoff}. In their works, they showed that a mass limit can be
also achieved in neutron stars. This mass limit, known as
Oppenheimer-Volkoff limit, appears when the neutron star pressure is 
sufficiently large, since it contributes to the mass-energy of the system 
turning the 
gravitational field large enough that it cannot be counteracted by the
pressure itself. It is worth mentioning that this limit, as well
as that the Chandrasekhar limit, have also been determined by Landau using
heuristic arguments, see \cite{landau}. In their works, Tolman \cite{tolman}
and Oppenheimer and Volkoff \cite{oppievolkoff} developed a consistent
method to describe a star in equilibrium configuration. This method is prone
to numerical integration. Once defined the matter distribution, they wrote
the gradient pressure in a very convenient form. This equation is known as
hydrostatic equilibrium equation or Tolman-Oppenheimer-Volkoff (TOV)
equation. To allow a complete description of the stars, all these
equilibrium configurations can be connected smoothly with the Schwarzschild
vacuum exterior solution.

The polytropic equation of state \eqref{EoS1} and the TOV equation of 
hydrostatic
equilibrium were used together for the first time by Tooper
\cite{tooper_poli}. He discussed the structure of polytropic stars
(polytropes) through the numerical integration of TOV equation. Despite 
that this equation of state describes spherical objects in a very simple manner, 
its use has some drawbacks. At very high pressures, it leads to obtain 
values of the sound speed higher than the speed of light, violating the
principle of causality. Thus, it is understood that a generalization of the
EoS $1$ is required. The most reasonable generalization of the polytropic
equation of state was determined by Tooper in \cite{tooper}. He showed that
the pressure and the energy density of the generalized polytropic equation
of state (EoS $2$) obey the relations
\begin{equation} \label{EoS2}
 \begin{split}
 &  p=\omega\delta^{\gamma},\\
 & \rho=\delta+p/(\gamma-1),
 \end{split}
\end{equation}
respectively, where $\delta$ represents the rest mass density and $\gamma$ is 
the polytropic exponent. Equations of state of this form
\eqref{EoS2} have been used to study neutron stars, in which the neutrons are
non-relativistic, and in white dwarfs, in which the electron gas is extremely
relativistic (see, e.g., \cite{tooper}). For white dwarf models, where the
fluid pressure is small in comparison to the energy density, EoS $2$ is
equivalent to EoS $1$, because in that situation we may neglect the
pressure in the second term on the right-hand-side of Eq.~\eqref{EoS2}, and
take $\delta\simeq \rho$. The
equilibrium configurations determined with the EoS $2$ are named by Thorne
as the relativistic polytropic models or relativistic polytropes (for short)
\cite{thorne_review}, henceforth, these names will be used throughout this
work. It is
important to mention that a brief comparison between 
non-relativistic polytropes and relativistic polytropes without and with 
cosmological constant have been considered respectively in \cite{hledik2005} 
(see also \cite{pandey1990,pandey1989}) and \cite{hledik2004}, and
in the presence of anisotropy in \cite{herrera2013_2}.

\subsection{Charged spheres and the TOV method}

 The first analyses on charged objects by means of the TOV method were
developed by Bekenstein in \cite{bekenstein}. He generalized the hydrostatic
equilibrium equation, i.e., the TOV equation, to include the effects
electric charges and electrostatic fields. From then on, different
works addressing the influence of electric charge in the 
structure of compact objects were reported. Among them, we find the studies
of the influence of the electric charge in the equilibrium configurations of
compact stars where the fluid follows the EoS $1$, e.g., see
\cite{zhang,raymalheirolemoszanchin,siffert,ALZ}. In
Refs.~\cite{raymalheirolemoszanchin,siffert} the authors focused on studying
the effects of the electric charge on the structure of compact cold stars.
In these works, the modified TOV equation was solved considering the
EoS $1$ with $\gamma=5/3$ and a charge density proportional 
to the energy density, $\rho_e=\alpha\rho$ ($\alpha$ being a constant that
obeys the constraint $0\leq\alpha\leq1$). Arba\~nil, Lemos and Zanchin (ALZ) in 
\cite{ALZ} also studied the structure of electrically charged objects 
considering the EoS $1$, for different $\gamma$, and with the charge 
distribution $\rho_e=\alpha\rho$. The authors found that extremely charged 
polytropic stars 
with $\gamma\rightarrow\infty$ are structures with the total charge $Q$ close 
to the total mass $M$, $Q\simeq M$, and the total radius $R$ close to the 
gravitational radius $R_+$, $R\simeq R_+\simeq M$. This indicates that the 
solutions are close to the quasiblack hole configurations, i.e., structures 
with $Q=M$ and $R=M$ and quasi-horizons (see, e.g., \cite{lz1}). All the 
aforementioned charged static
equilibrium configurations are matched smoothly with the Reissner-Nordstr\"om 
vacuum exterior solution.

\subsection{Compactness bounds and quasiblack hole configurations}

The solutions of compact objects found in general relativity are connected
with the Buchdahl bound \cite{buchdahl}. This bound states that the radius
$R$ and the gravitational mass $M$ of a sphere of perfect fluid in
hydrostatic equilibrium, in which the energy density is non-increasing
outward, satisfies the inequality $R/M\geq 9/4$. If a star shrinks to a size
that violates this bound, it eventually turns into a black hole. This bound
is saturated by the interior Schwarzschild solution in the limit
of infinite central pressure \cite{incom_schwarzschild} (see also
\cite{ALZ2014}).  This is the Schwarzschild interior limit,
which saturates the Buchdahl bound in the sense that an incompressible fluid
with an infinite central pressure gives the upper limit of the bound, $R/M =
9/4$. The Buchdahl bound is a general result, i.e., it is
independent of the equation of state used.

The charged static equilibrium solutions found in an Einstein-Maxwell system
are related with the Buchdahl bound for the electric case
\cite{andreasson_charged}, i.e., with the Buchdahl-Andr\'easson bound, in
which the hydrostatic equilibrium configuration 
satisfies the condition $R/M\geq9/\left(1+\sqrt{1+3Q^2/R^2}\right)^{2}$.
When $Q=R$, the Buchdahl-Andr\'easson bound is saturated, i.e., we obtain
$R=M$ and also $Q=M$. In other words, this bound is 
saturated by a quasiblack hole configuration. As shown in
Ref.~\cite{lemos_zanchin_Guilfoyle2015}, the Buchdahl-Andr\'easson bound is
saturated by the Guilfoyle solutions \cite{guilfoyle} for charged spheres in
the limit where the central pressure attains arbitrarily large values, in
full analogy to the Schwarzschild interior limit. As far as we know, this
is the only solutions that saturates such a bound. As verified
in Refs.~\cite{ALZ,ALZ2014}, charged fluids satisfying the 
non-relativistic polytropic equation of state and a charged incompressible 
fluid do not saturate the Buchdahl-Andr\'easson bound.

It is important to stress that the quasiblack hole limit is found using 
different equations of state and different distributions of electric charge. 
Such limiting solutions have been found, e.g, for an incompressible fluid, 
i.e., $\rho=$ constant, with a distribution of electric charge which follows a 
particular function of the radial coordinate 
\cite{defelice_siming,defelice_yu,anninos_rothman}, and when the charge 
density is proportional to the energy density \cite{ALZ,ALZ2014}. They are 
also obtained in works that use an equation of state for electrically charged 
dust, i.e., $p=0$ \cite{bonnor_wick,lemosweinberg}. These objects are also 
obtained in 
\cite{lemosezanchin_QBH_pressure,lemos_zanchin_Guilfoyle2015,
lemosezanchin_plethora} where it is considered the Cooperstock-De la 
Cruz-Florides equation of state \cite{cooperstock1978,florides1983, 
guilfoyle}. The general properties of quasiblack holes are defined in 
\cite{lz1,lz4}.

\subsection{This work}

We are interested in comparing equilibrium configurations of charged fluid
spheres in the presence or absence of electric charge, obtained from the
two equations of state cited above, namely the non-relativistic polytropic 
equation \eqref{EoS1} and the relativistic polytropic equation \eqref{EoS2}.
For short, we refer to the respective configurations as
non-relativistic polytropic stars (or non-relativistic polytropes), and 
relativistic polytropic stars (or relativistic polytropes).
Very compressed
objects and the compactness bounds and quasiblack hole limits are the main
objects of interest here. Let us mention once again that the major part of
the analysis in the case of the non-relativistic polytropic equation of
state was performed in Ref.~\cite{ALZ}. The main aim now is the relativistic
polytropic spheres, and the comparison to the non-relativistic polytropic 
equation is also done here.
For the sake of comparison with previous works, the distribution of electric
charge in the structure of the star is assumed
to follow the equation $\rho_e=\alpha\rho$. Some features mentioned
previously are investigated in this paper. For these objects we study the
Chandrasekhar limit, the Oppenheimer-Volkoff limit, the Buchdahl bound, the
Buchdahl-Andr\'easson bound, and the quasiblack hole limit. The speed of 
sound throughout a given sphere and the redshift at the surface of the sphere 
are also investigated.

The article is structured according as follows. 
In Sec.~\ref{sec-basicequations} we write the TOV equation with the inclusion 
of the electric charge. To complete the set 
of equations, we also present the equations of state to be used, as well as
the charge density profile and the boundary conditions.
Sec.~\ref{results1} is dedicated to compare the structure of charged
non-relativistic polytropes with the charged relativistic polytropes for
different values of polytropic exponent $\gamma$. We analyze the
Chandrasekhar limit, the Oppenheimer-Volkoff limit, the Buchdahl bound and
the Buchdahl-Andr\'easson bound.
We present the dependence of the mass, the radius, and the charge of the
charged spheres as a function of the polytropic exponent. We also present
the dependence of the mass, radius and charge against the charge fraction.
Some physical properties of the fluid for an arbitrarily large polytropic
exponent $\gamma$ are given in Sec.~\ref{sect-largegamma}. The dependence 
of the speed of sound as a function the polytropic exponent is accomplished 
in Sec.~\ref{sound-section}. Section~\ref{qbh-section} is devoted to study the 
quasiblack hole limit and the redshift on the surface of a quasiblack hole. 
In every section we present the new results for the relativistic polytropic 
spheres and, for a better comparison, the results of the non-relativistic 
polytropic spheres are also reviewed.
In Sec.~\ref{sec-conclusion} we conclude.

Finally, it is worth mentioning that, unless otherwise
stated, geometric units shall be used throughout the text, so that $c=1=G$.

\section{General relativistic charged perfect fluid}
\label{sec-basicequations}

\subsection{Equations of structure}

With the purpose of analyzing the properties of static charged perfect
fluid distributions we take the line element, in Schwarzschild 
coordinates, as
\begin{equation}
\label{geral_metric}
ds^2=-B(r)dt^{2}+A(r)dr^2+r^{2}d\theta^2+r^{2}\sin^{2}\theta d\phi^{2} ,
\end{equation}
with the metric potentials $B(r)$ and $A(r)$ depending on the radial
coordinate $r$ only.
The Einstein-Maxwell equations furnish the following non-identically zero
equations
\begin{equation}\label{continuidad da carga}
\frac{dq(r)}{dr}=4\pi\rho_{e}(r)\,r^{2} \sqrt{A(r)},
\end{equation}
and
\begin{eqnarray}
\frac{1}{A(r)}\left[1-\frac{r}{A(r)}\frac{dA(r)}{dr}\right]\!&=&\!
   1- 8\pi\, r^2\left[\rho(r) +\frac{q^{2}(r)}{8\pi r^{4}}\right]\!,  
\label{G00final12} \;\;\;\\
\frac{1}{A(r)}\left[1+\frac{r}{B(r)}\frac{dB(r)}{dr}\right]\!&=&\!
   {1} + 8\pi r^2\left[p(r)-\frac{q^{2}(r)}{8\pi r^{4}}\right]\!. \;\;
\label{G11final12}
\end{eqnarray}
Function $q(r)$ represents the electric charge within a sphere of radius
$r$, $\rho_e(r)$ is the electric charge density, and $\rho(r)$ and $p(r)$ 
stand respectively for the energy density and the pressure of the fluid.

We now introduce the mass function $m(r)$ through the relation
\begin{equation}\label{funcion metrica}
A^{-1}(r)=1-\frac{2m(r)}{r}+\frac{q^{2}(r)}{r^{2}}.
\end{equation}
Considering this mass function \eqref{funcion metrica}, we have that
Eq.~(\ref{G00final12}) can be written  in the form
\begin{equation}\label{continuidad de la masa}
\frac{dm(r)}{dr}=4\pi\rho(r)
r^{2}+\frac{q(r)}{r}\left[\frac{dq(r)}{dr} \right].
\end{equation}
This differential equation represents the continuity equation,
i.e., the mass-energy conservation. 

An additional relation may be obtained from the Bianchi identity
($\nabla_{\mu}T^{\mu\nu}=0$)  which, with metric \eqref{geral_metric},
yields
\begin{equation}  \label{conservacion2}
\frac{dB(r)}{dr}=\frac{2\,B(r)}{p(r)+\rho(r)}\left[\frac{q(r)}{4\pi
r^{4}}\frac{ dq(r)}{dr}-\dfrac{dp(r)}{dr} \right].
\end{equation}
Replacing Eqs.~(\ref{continuidad da carga}) and (\ref{G11final12}) into
Eq.~(\ref{conservacion2}) we obtain the modified TOV equation with the
inclusion of electric charge \cite{bekenstein},
\begin{equation}\label{tov}
\frac{dp}{dr}=-(p+\rho)A{\left(4\pi p\,r+\frac{m}{r^{2}} 
-\frac{q^{2}} {r^{3}}\right)}+\rho_{e} \sqrt{{A}}\,\dfrac{q}{r^{2}},
\end{equation}
where, with the purpose of simplifying the equation, 
the explicit dependence of the variables on the radial coordinate was removed, 
i.e., we have written
$p(r)=p$, $\rho(r)=\rho$, $\rho_e(r)=\rho_e$, $m(r)=m$, $q(r)=q$, and $A(r)=A$.
Taking $q=0$ in this equation, the original TOV~\cite{oppievolkoff,tolman}
equation is recovered.

\subsection{The equation of state and charge density profile}
\label{EOS_charge_DP}

In order to look for equilibrium solutions, it is necessary to solve 
simultaneously equations (\ref{continuidad da carga}), (\ref{funcion 
metrica}), (\ref{continuidad de la masa}), and (\ref{tov}). These four 
equations contain six variables $q(r)$, $A(r)$, $m(r)$, $\rho(r)$, 
$p(r)$, and $\rho_{e}(r)$, forming an incomplete set of equations. To
complete the system, as usual, an equation of
state relating the pressure with the energy density and, for
the charged fluid, a relation defining the charge density profile are
supplemented. Once closed the system of equations, together with the
boundary conditions imposed in the system, these are solved 
numerically.

As stated earlier, we employ two different polytropic equations of state to
connect the fluid pressure and the fluid energy density.  
In the following, the equation of state \eqref{EoS1} (EoS $1$)
and its respective results are called case $1$, while the equation 
\eqref{EoS2} (EoS $2$) and its respective results are called case $2$. 

As in previous works, e.g.,
\cite{zhang,raymalheirolemoszanchin,siffert,ALZ,arbanil_malheiro, 
lemos_lopes_quinta_zanchin2015}, the electric charge distribution is
considered proportional to the energy density, as follows 
\begin{equation}\label{densicarga_densimasa}
\rho_e=\alpha\rho,
\end{equation}
where $\alpha$ is a dimensionless constant that we call the charge
fraction, which is constrained to the interval $\alpha \in 
\left[0,\,1\right)$.

\subsection{The boundary conditions and the exterior line element}

The complete set of differential equations are provided with a set of
additional constraints, allowing to find the equilibrium solutions. For the
non-relativistic polytropic fluids,case 1, the conditions at the center of 
the spheres are $q(r=0)=0$, $m(r=0)=0$, and $\rho(r=0)=\rho_{c}$. 
For the relativistic polytropic fluids, case 2,
the conditions at the center of the spheres are $q(r=0)=0$, $m (r=0)=0$,
and $\delta(r=0)=\delta_{c}$.
In both cases, the surface of the objects is found when $p(r=R)=0$. The input 
data for the numerical calculation are the central energy density $\rho_{c}$
and the central rest mass density $\delta_{c}$, respectively, for case 1 
and case 2, the polytropic constant $\omega$, the polytropic exponent 
$\gamma$, and the charge fraction $\alpha$. 

In both cases, the interior solution connects smoothly with the exterior 
solution given by the Reissner-Nordstr\"om metric
\begin{equation}
ds^2 = - F(r) dT^2 + \frac{dr^2}{F(r)}+
r^2\left(d\theta^2+\sin^{2}\theta d\phi^{2}\right),
\end{equation}
with $F(r) = 1 -2M/r + Q^2/r^2$. The total mass and the total charge of
the sphere are represented by $M$ and $Q$, respectively. The
time $T$ is proportional to the inner time $t$, and the radial
coordinate $r$ is identical to the interior region. The full set boundary
conditions at the surface of the sphere are $p(R)=0$
(this condition is used to determine the radius of the star), $m(R)=M$,
$q(R)=Q$, and the continuity of the metric functions $B(R)=1/A(R)=F(R)$.

\section{The structure of relativistic charged polytropic spheres}
\label{results1}

\subsection{General remarks}

Here the structure of charged spheres is analyzed for 
different values of the
exponent $\gamma$, and for different values of $\rho_c$ and $\delta_c$,
respectively, in the case $1$ and in the case $2$. In order to make a proper
comparison of our results with those found in the literature, 
some similar considerations have to be made, both for the choice of the
polytropic constant as for the normalization used in the numerical
integration. Therefore, for the two equations of state used, following 
Ref.~\cite{ALZ}, we take the polytropic constant as
$\omega=1.47518\times 10^{-3} \left(1.78266\times
10^{15}\mathrm{kg/m^{3}}\right)^{1-\gamma}$. 
For such a choice, the normalization factor adopted during the numerical
integration of the TOV  equation is $\rho_0=1.78266\times10^{15}[\rm
kg/m^3]$ for the case $1$, and 
$\delta_{0}^{*}=1.78266\times10^{15}[\rm kg/m^3]$ for the case $2$. Then,
during the numerical calculations, the polytropic constant is written as
\begin{equation}\label{omega}
 \omega = \bar\omega\, {\rho_0}^{1-\gamma}, \quad \omega =
\bar\omega\,{\delta_0^*}^{1-\gamma},
\end{equation}
respectively, for the EoS $1$ and EoS $2$. The dimensionless polytropic
constant is now $\bar\omega = 1.47518\times 10^{-3}$.

For the numerical solutions, the equations of structure \eqref{continuidad
da carga}, \eqref{funcion metrica}, \eqref{continuidad de la masa}, and
\eqref{tov}, the equation of state, the charge density profile, and the
boundary conditions are written in a dimensionless form. 
For the non-relativistic polytropic case, Eq.~\eqref{EoS1}, this was
done in Ref.~\cite{ALZ}, while for the relativistic polytropic case,
Eq.~\eqref{EoS2}, the normalized equations are shown 
in Appendix \ref{eq_adimen_politropico}. 
Once the values of $\alpha$, $\gamma$, and $\rho_{c}$ (or $\delta_c$,
depending on the case) has been fixed, the system of equations are solved
numerically  by using the fourth-order Runge-Kutta method.

For convenience of the numerical analysis we restore the
gravitational constant, $G =7.42611\times10^{-28} [{\rm m/kg}]$, but keep
the speed of light $c = 1$.

Along this section, we compare the equilibrium configurations of charged
relativistic polytropes to the charged non-relativistic polytropes
studied in Ref.~\cite{ALZ}. Additionally, some particular limits, such as 
the Chandrasekhar limit, the Oppenheimer-Volkoff limit, the Buchdahl bound,
the  Buchdahl-Andr\'easson bound, and the quasiblack hole limit are
tested for the new equation of state (case 2). 
To study the Chandrasekhar and Oppenheimer-Volkoff limits, the
polytropic exponent is fixed and the interval of both the
central energy density and the central rest mass density is varied from
$10^{13}[\rm kg/m^3]$ to $10^{20}[\rm kg/m^3]$. In turn, to study the 
the Buchdahl and the Buchdahl-Andr\'easson bounds and quasiblack hole
limit, we fixed the central energy density in $10\,\rho_0$ in case $1$
and the central rest mass density $10\,\delta_0^{*}$ in the case $2$. For
those densities, the largest value of the polytropic
exponent that produces good numerical results is $17.0667$ for the case
$1$, and $17.1109$ for the case $2$.
Thus, in order to realize a comparison between the results found for the
two equations of state, case $1$ \eqref{EoS1} and case 2 \eqref{EoS2}, we
take the values of the polytropic exponent in the same range
$4/3\leq\gamma\leq17.0667$.

\subsection{The radius against the mass for fixed polytropic
exponent: The Chandrasekhar and the Oppenheimer-Volkoff limits} 

\begin{figure}[ht]
\centering
\includegraphics[scale=0.29]{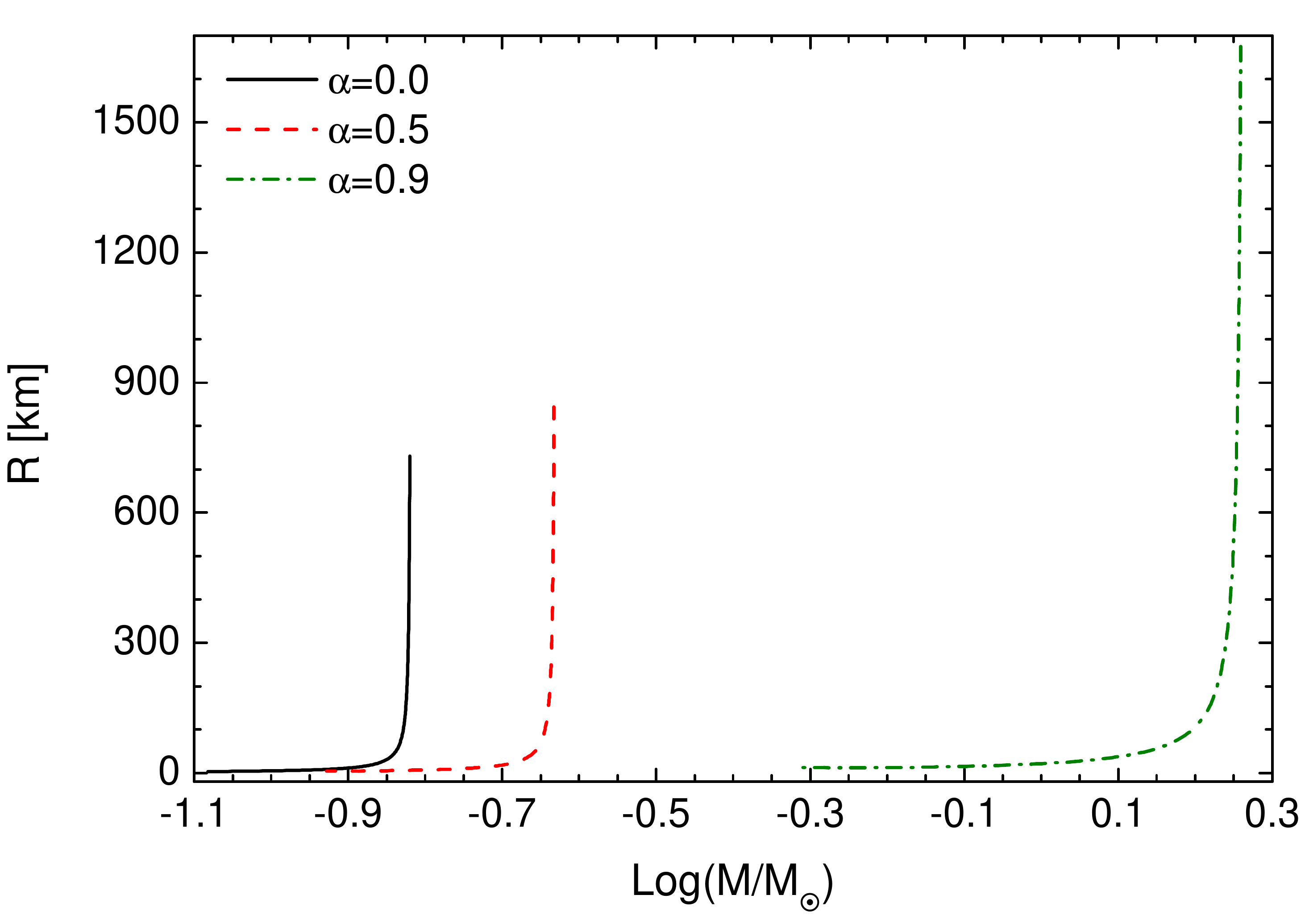}
\caption{The radius of the sphere against the mass for $\gamma=4/3$ and three
values of charge fractions, as indicated. The top (bottom) panel is for
the EoS $1$ (EoS $2$). The central energy density and the 
central rest mass density, respectively, are in the 
interval $[ 10^{13}({\rm kg/m^3}),\,  10^{20}({\rm kg/m^3})]$. In both
cases the Chandrasekhar and Oppenheimer-Volkoff mass limits are found.
Notice that these limits do not depend on the equation of state.}
\label{RM_4_3}
\end{figure}

\begin{figure}[ht]
\centering
\includegraphics[scale=0.29]{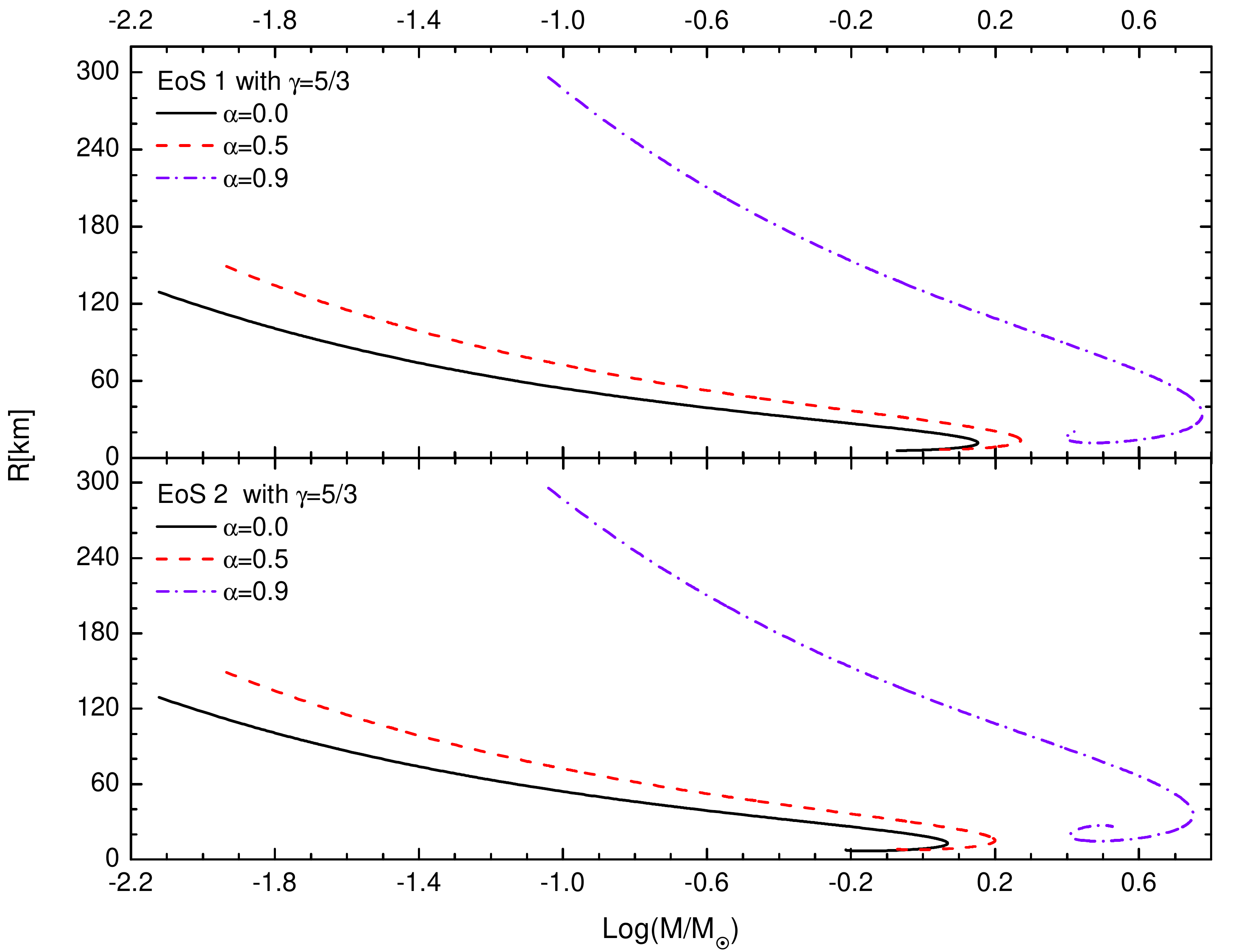}
\caption{The radius versus the mass of the spheres for the polytropic
exponent  $\gamma=5/3$ and a few values of charge fractions, as indicated.
The top (bottom) panel is for the EoS $1$ (EoS $2$). The central energy
density and the central rest mass density, respectively, are in the 
interval $[ 10^{13}({\rm kg/m^3}),\,  10^{20}({\rm kg/m^3})]$.  The 
Oppenheimer-Volkoff mass limit appears clearly, however, there is no  
Chandrasekhar mass limit for $\gamma\neq4/3$.}
\label{RM_5_3}
\end{figure}

Let us start investigating the behavior of the radius and the mass of the
charged fluid spheres for different central densities. 
Figs.~\ref{RM_4_3} and \ref{RM_5_3} contain the
curves for the radius as a function of the mass of the spheres (normalized
to the Sun's mass $M_{\odot}$) for $\gamma=4/3$ and $\gamma=5/3$,
respectively, and three charge fraction values $\alpha=0.0$, $0.5$ and
$0.9$. The upper panel in each figure shows the results for the
non-relativistic polytropes (case 1), while the lower panel contains the 
results determined for relativistic polytropes (case 2). The central energy 
density and the central rest mass density are both varied from $10^{13}[{\rm 
kg/m^3}]$ to $10^{20}[{\rm kg/m^3}]$.

In Fig.~\ref{RM_4_3}, for $\gamma=4/3$, the curves in top panel, case $1$,  
indicate that the mass and the radius of the spheres decrease with increasing 
of the central energy density. Similar behavior is shown by the curves in the 
bottom panel, case $2$, showing that the mass and the radius of the 
relativistic polytropes decrease with the increase of the central
rest mass density. Moreover, the radius is an increasing
function of the mass, the smaller values of the energy (rest mass) density
correspond to the higher values of $R(M)$, on the right end of each
curve. In addition, we note that the Chandrasekhar limit is found at zero
radius, and the Oppenheimer-Volkoff limit appears at the point where the
vertical lines turn to the left. It is clear the influence of the electric
charge. For $\alpha=0.9$ the mass of the stars for the same central
density are about three times larger than for $\alpha=0.0$. The radius, of
course, also grows with the charge fraction approximately at the same rate
as the mass, almost independently of the equation of state.

In Fig~\ref{RM_5_3}, for $\gamma=5/3$, the curves in the top panel, case 
$1$, show that the radius of the spheres decreases with the mass, while the 
central energy density grows.
Similar behavior is shown by the curves in the bottom panel, case 2, the 
radius of the relativistic polytropic spheres decreases with the mass, while 
the central rest mass density grows.
For high values of $\rho_c$ and $\delta_c$ we observe that the curves in
both panels present a spiraling behavior. When $\gamma=5/3$ 
only the Oppenheimer-Volkoff limit appears in the point where the inclined 
lines are folded to the left. For $\gamma\neq4/3$, the  
Chandrasekhar limit does not appear. The radius and mass grow with the charge 
factor as for other polytropic exponents, the growth rate being approximately 
independent of the equation of state chosen. 

It is worth mentioning that, irrespective the $\alpha$ used in 
Figs.~\ref{RM_4_3} and \ref{RM_5_3}, we find that the masses of 
(non-relativistic) polytropes and relativistic polytropes are very close to 
each other for low and equal values of $\rho_c$ and $\delta_c$. For instance, 
in the case of $\gamma=4/3$, $\alpha=0.9$, and $\rho_c=\delta_c=10^{13}[\rm 
kg/m^3]$ we obtain a mass of $1.8173\,M_{\odot}$ in the case $1$, and a mass 
of $1.8154\,M_{\odot}$  in the case $2$. Independently of the $\alpha$ 
employed, the difference of these masses becomes more apparent when the value 
of $\rho_c=\delta_c$ is incremented. For example, still in the case for 
$\gamma=4/3$,  $\alpha=0.9$, but now with $\rho_c=\delta_c=10^{20}\,[{\rm 
kg/m^3}]$, the mass of the sphere in the case $1$ is $0.57874\,M_{\odot}$, and 
in the case $2$ it is $0.53214\,M_{\odot}$, a difference of about $19\%$. From 
this result we understand that the EoS $1$ and EoS $2$ are not equivalent for 
large values of energy (rest mass) density, as expected.

\subsection{The structure dependence of the relativistic charged spheres on
the polytropic exponent}

\subsubsection{Mass of the spheres against the polytropic exponent}

\begin{figure}[ht]
\centering
\includegraphics[scale=0.29]{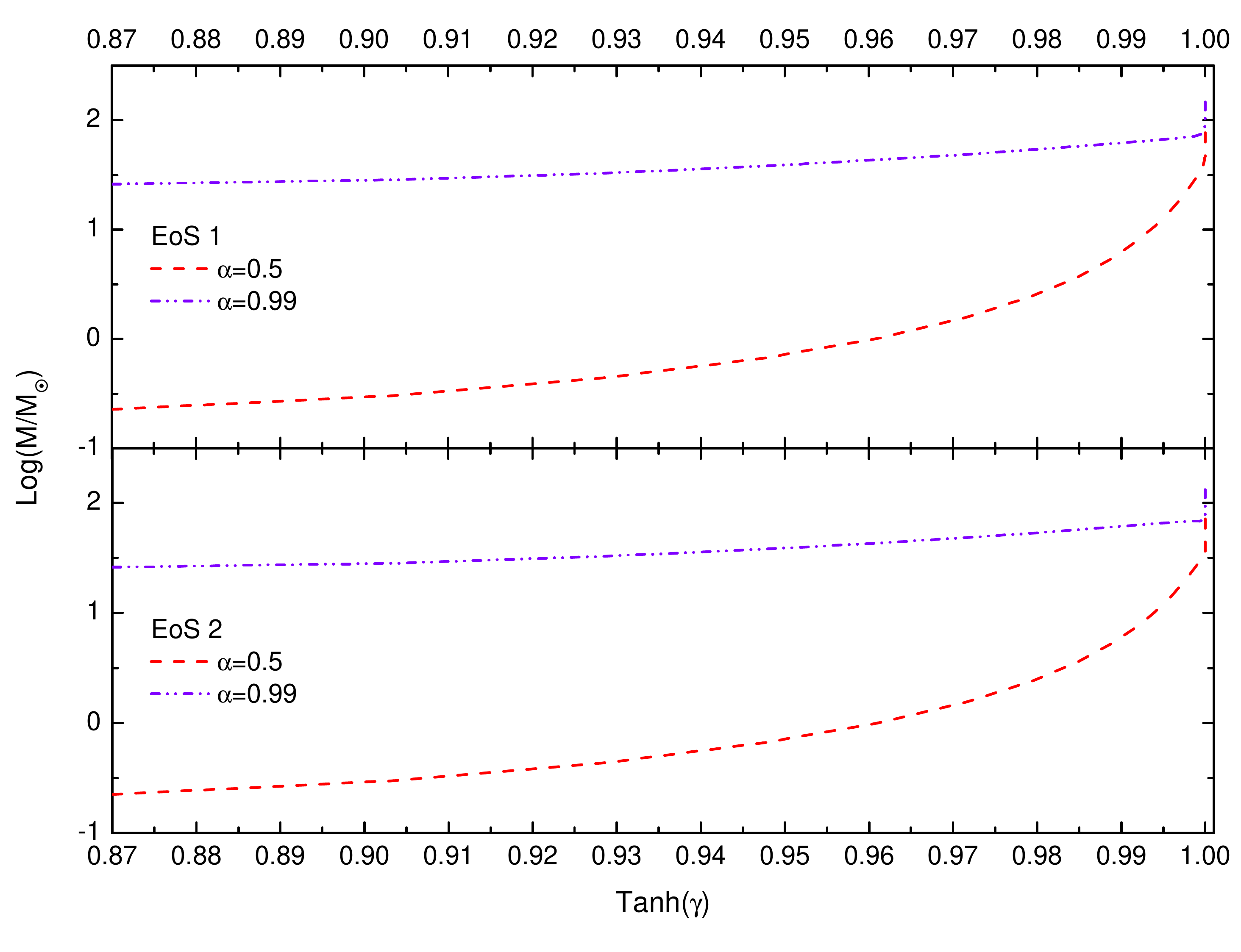}
\caption{Mass of the star as a function of the polytropic exponent $\protect
\gamma$ for two charge fractions, $\alpha=0.5$ and $\alpha=0.99$. 
The results for the EoS $1$, with the central energy density 
$\rho_c=1.78266\times10^{16}[\mathrm{kg/m^{3}}]$ 
are shown in the top panel, while the results for the EoS $2$, with the 
central rest mass density 
$\delta_{c}=1.78266\times10^{16}[\mathrm{kg/m^{3}}]$, are shown in the bottom 
panel.}
\label{Tanh_gam_M_Msol_EOSI_II}
\end{figure}

The numerical results obtained for the mass of the charged fluid spheres as a 
function of polytropic exponent $\gamma$ produce the graphs of Fig.
\ref{Tanh_gam_M_Msol_EOSI_II}.
The behavior of the ratio $M/M_\odot$ for some values of the charge fraction
$\alpha$ and for the two different equations of state under consideration is 
seen in that figure. 
As in the case of Figs.~\ref{RM_4_3} and \ref{RM_5_3}, the top panel is
for the EoS $1$ (case 1), Eq.~\eqref{EoS1}, and with the central energy 
density $\rho_c=1.78266\times10^{16}[\mathrm{kg/m^{3}}]$, and the bottom
panel is for the EoS $2$ (case 2), Eq.~\eqref{EoS2}, and with the central 
rest mass density $\delta_c=1.78266\times10^{16}[\mathrm{kg/m^{3}}]$.
The polytropic exponent considered in both cases is in the interval
$4/3\leq\gamma\leq17.0667$.
For low values of $\gamma$, the masses found in both cases are 
very close to each other. This result is expected
since, in this regime, i.e., taking into account that the central energy
density is not very high, the relativistic effects on the equation of state
for the fluid are small and EoS $1$ and EoS $2$ are equivalent.
In both cases, we observe that the mass increases very fast 
with the polytropic exponent. For instance, analyzing the mass in the points
$\gamma=4/3$ and $17.0667$ for $\alpha=0.5$, we obtain that it grows
approximately $36,431\%$ in case $1$,  and around $34,242\%$ in case $2$. In 
turn, for $\alpha=0.99$ we obtain that the mass grows at about $488\%$ 
for EoS $1$, and almost $456\%$ for EoS $2$. The growth of the mass with the 
polytropic exponent $\gamma$ is explained in the same way for both equations 
of state, since a larger central pressure $p_c$ is obtained with a higher
$\gamma$.  In both cases, EoS $1$ and EoS $2$, the mass also
grows with the increase of charge fraction (see, also, Fig.
\ref{alpha_M_Msol_EOSI_II}). Again we note only a small difference
between the results of the two equations of state. The masses of the 
non-relativistic polytropes (case $1$) are of the order of $10\%$ larger 
than the masses of the relativistic stars (case $2$).

\subsubsection{Radius of the relativistic spheres against the polytropic
exponent}\label{radius_mass_sec}

\begin{figure}[ht]
\centering
\includegraphics[scale=0.29]{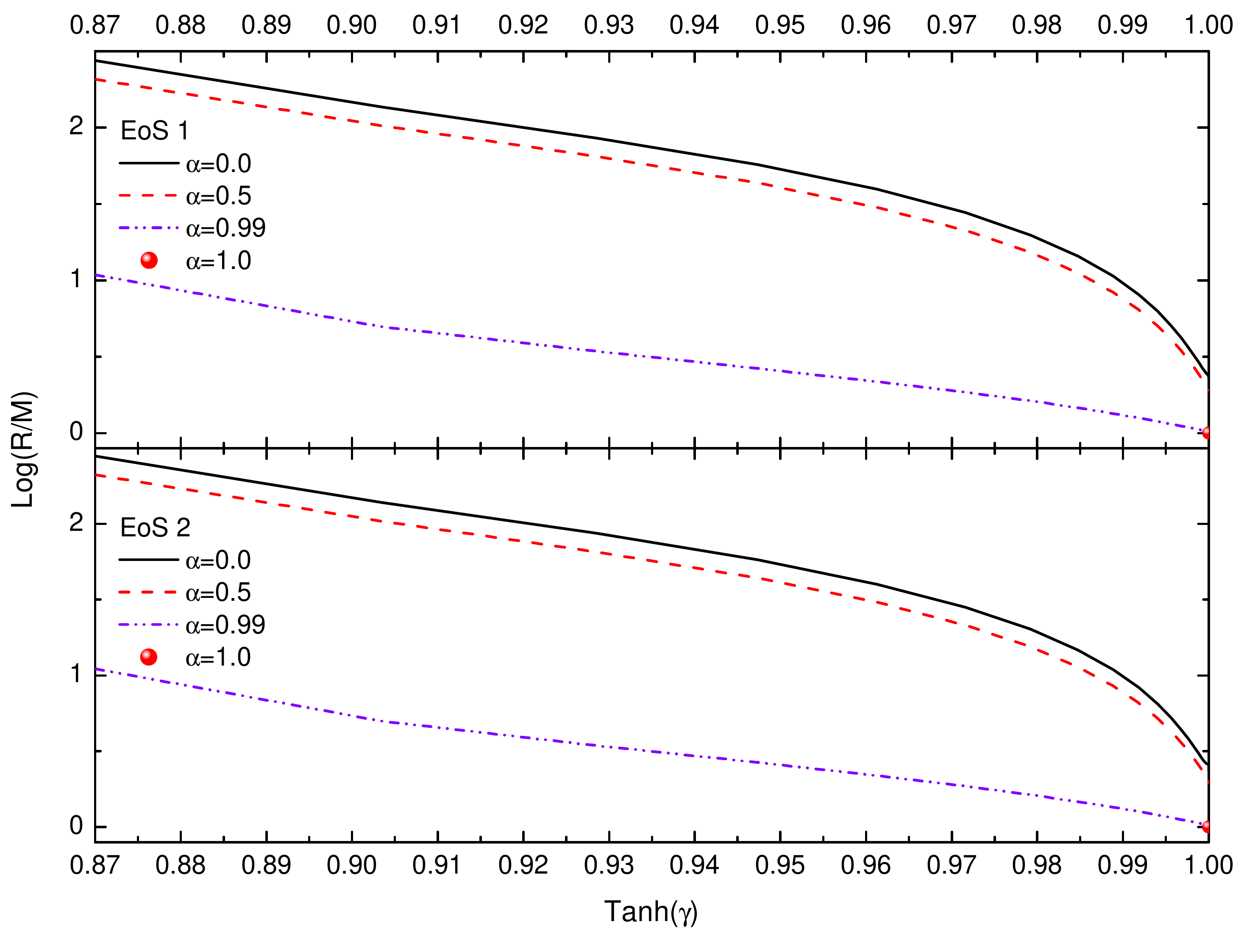}
\caption{Behavior of the ratio $R/M$ as a function of the polytropic exponent 
$\gamma$ for some values of the charge fraction. The top panel is for the EoS 
$1$ with the central energy density 
$\protect\rho_c=1.78266\times10^{16}[\mathrm{kg/m^{3}}]$. The bottom panel 
is for the EoS $2$ with the central rest mass density $
\protect\delta_c=1.78266\times10^{16}[\mathrm{kg/m^{3}}]$.}
\label{Tanh_gam_R_M_EOSI_II}
\end{figure}

The radius to mass ratio of the sphere against the polytropic exponent is 
shown in Fig. \ref{Tanh_gam_R_M_EOSI_II}, where we plot the ratio $R/M$ versus 
$\gamma$ for two values of charge fraction, $\alpha=0.5$ and $0.99$, and for 
the two equations of state. The results for the EoS $1$ with the central 
energy density $\rho_c=1.78266\times10^{16}[\mathrm{kg/m^{3}}]$ are shown in 
the top panel, while the results for EoS $2$ with the central rest mass 
density $\delta_c=1.78266 \times10^{16}[\mathrm{kg/m^{3}}]$ are shown in the 
bottom panel. As in the other figures of the present section, the polytropic 
exponent values are between $4/3$ and $17.0667$.

It can be observed in Fig. \ref{Tanh_gam_R_M_EOSI_II} that the ratio $R/M$ 
decreases with the increment of $\gamma$, reaching its minimum value at the 
maximum value of the polytropic exponent $\gamma=17.0667$. In the uncharged 
case, $\alpha=0.0$, we see that the minimum value of the radius to mass ratio 
is approximately $R/M=2.279$ in the EoS $1$ case, and $R/M=2.352$ in the EoS 
$2$ case. From these results we understand that if we extrapolate the 
polytropic exponent $\gamma$ to infinity, so to reach the incompressible fluid 
configuration, the Buchdahl bound \cite{buchdahl} is saturated in the case 
$1$. However in the EoS $2$ case the upper limit of the Buchdahl bound is not 
attained (see also Refs. \cite{ALZ,ALZ2014}). The different result found in 
the EoS $2$ case may be explained by observing that the effects of a large 
(infinite) central pressure is counterbalanced by the effects of a large 
(infinite) energy density and, as consequence, by a large attractive 
gravitational force, preventing the object to reach the maximum compactness 
set by the Buchdahl bound. The main point that may explain this different 
degree of compactness is that the central energy density is finite in case 
$1$, while it diverges in case $2$. Notwithstanding, in the extremely charged 
case, $\alpha=0.99$, we have $R/M\simeq1.027$ for the case EoS $1$, and 
$R/M\simeq1.025$ for the case EoS $2$. These values of $R/M$ are close to the 
maximum compactness of a charged object, $R/M=1.0$. From this we understand 
that for large (infinite) values of $\gamma$ the Buchdahl-Andr\'easson bound 
\cite{andreasson_charged} is saturated in the limit $\alpha \rightarrow 1$.

\begin{figure}[ht]
\centering
\includegraphics[scale=0.29]{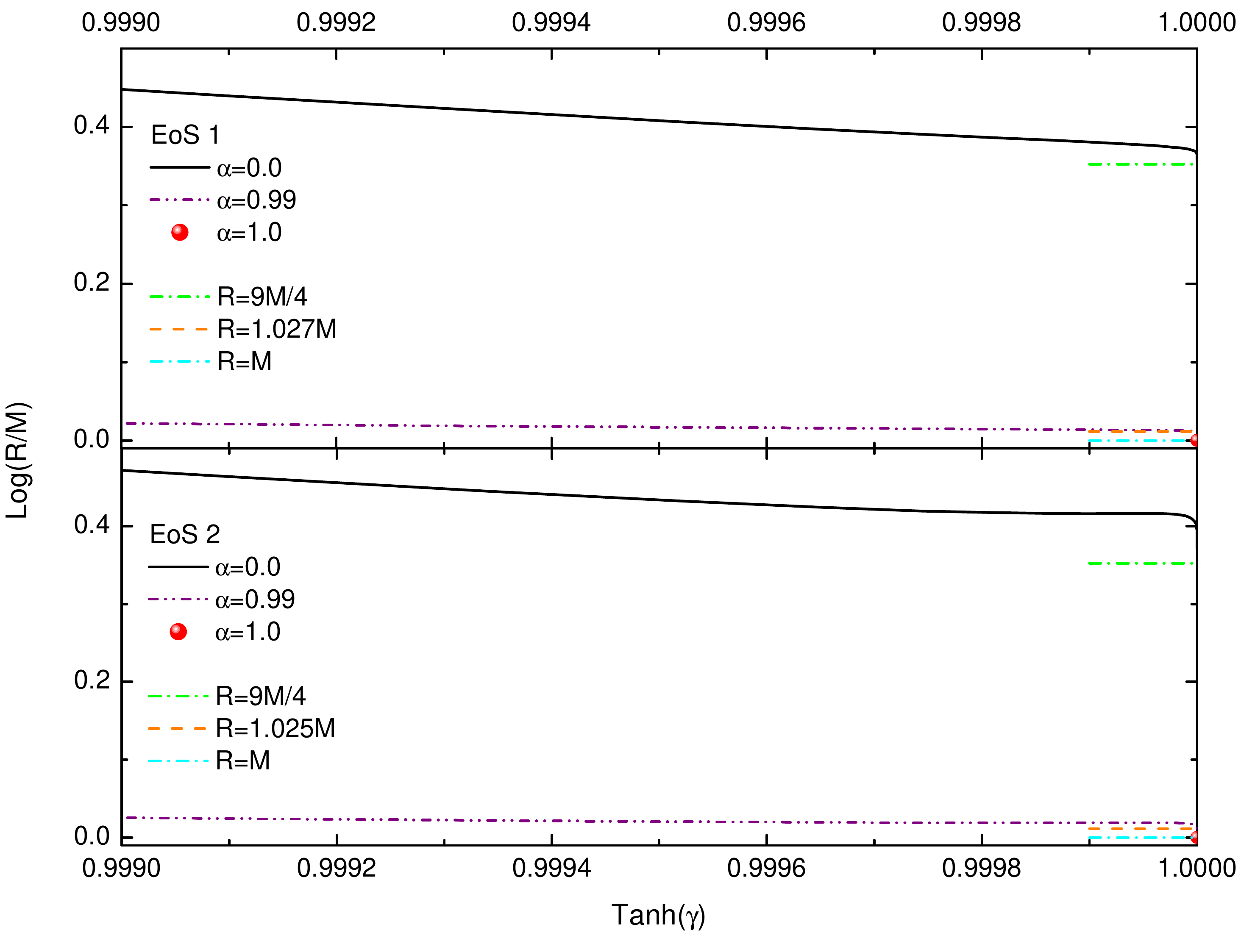}
\caption{The ratio $R/M$ against the polytropic exponent. This is an 
amplification of the region of Fig~\ref{Tanh_gam_R_M_EOSI_II} corresponding 
to high values of $\gamma$. The data are the same as in that figure. }
\label{Tanh_gam_R_M_EOSI_II_ampliado}
\end{figure}

For a better visualization of the results shown in 
Fig.~\ref{Tanh_gam_R_M_EOSI_II}, the relation $R/M$ for the highest values of 
$\gamma$ we have obtained are shown in the Fig. 
\ref{Tanh_gam_R_M_EOSI_II_ampliado}. From this figure it can be seen in detail 
the extreme limits for $\alpha=0.0$ as well as for $\alpha=0.99$. For the case 
without charge, in the top panel, unlike what is shown in the bottom panel, we 
see that the Buchdahl limit is close to be attained. In turn, for 
$\alpha=0.99$, on the top as well as in the bottom panel, we see that the 
Buchdahl-Andr\'easson bound is close to be saturated (see 
Sect.~\ref{sect-largegamma} for more details).

\subsubsection{Charge of the spheres against the polytropic exponent}
\label{charge_mass_sec}

\begin{figure}[ht]
\centering
\includegraphics[scale=0.29]{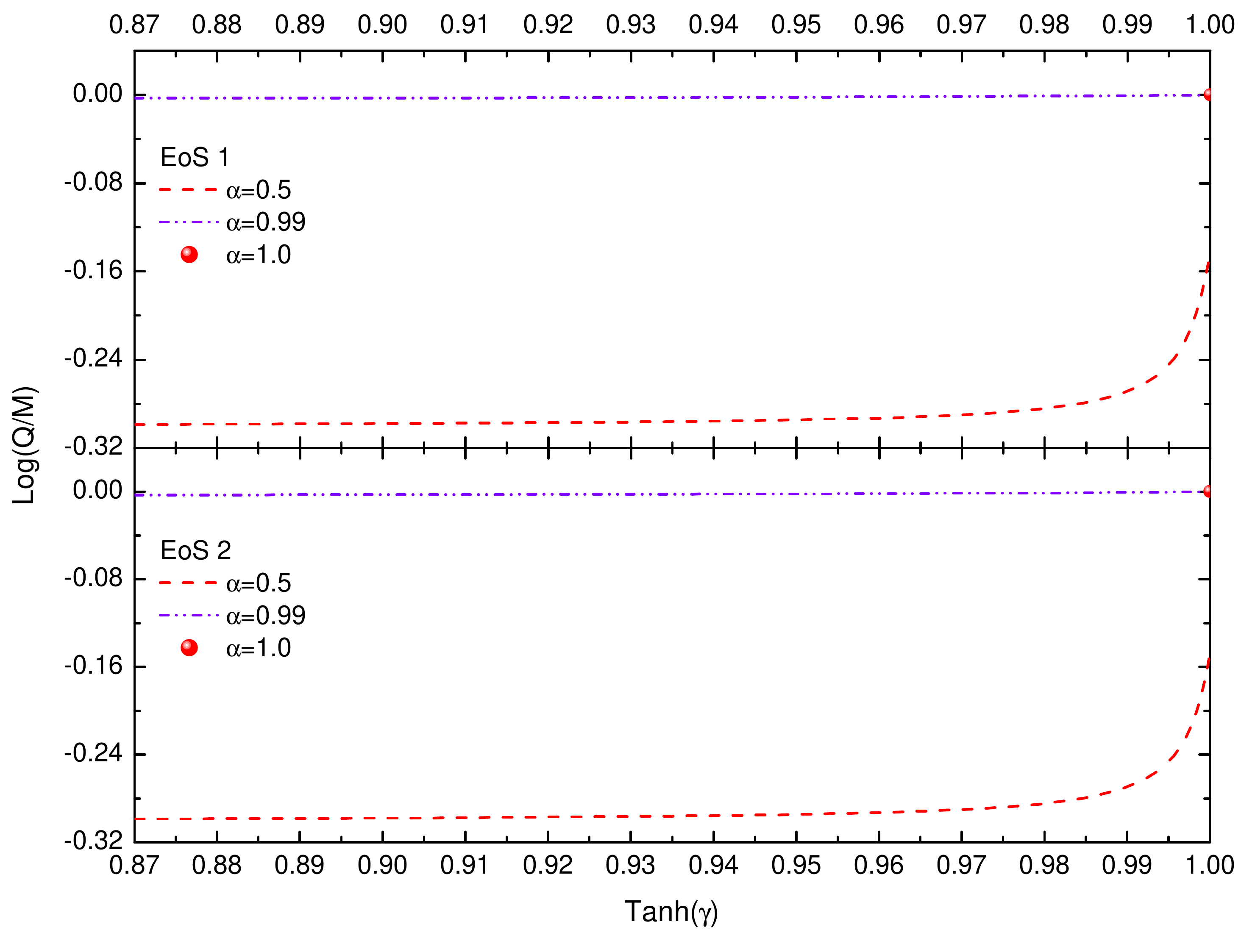}
\caption{The ratio $Q/M$ as a function of $\gamma$ for two values of
the charge fraction, $0.5$ and $0.99$. The top panel is for
the polytropic equation of state (EoS $1$) with
$\rho_c=1.78266\times10^{16}[\mathrm{kg/m^{3}}]$, while the bottom panel is
for the relativistic polytropic equation of state (EoS $2$) with 
$\delta_c=1.78266\times10^{16}[\mathrm{kg/m^{3}}]$.}
\label{Tanh_gam_Q_M_EOSI_II}
\end{figure}

The charge to mass ratio ($Q/M$) as a function of the polytropic exponent for 
two values  of charge fraction, $\alpha =0.5$ and $\alpha=0.99$, is plotted in 
Fig.  \ref{Tanh_gam_Q_M_EOSI_II}. As in the previous figures,  the curves in 
the top panel are obtained for the EoS $1$ with 
$\rho_c=1.78266\times10^{16}[\mathrm{kg/m^{3}}]$, while the curves in the 
bottom panel are for the EoS $2$ with 
$\delta_c=1.78266\times10^{16}[\mathrm{kg/m^{3}}]$. The behavior of the curves 
indicate that the relation $Q/M$ grows with $\gamma$ and $\alpha$, and it is 
essentially the same for the two equations of state. Note that in the extreme 
case, with $\alpha=0.99$ and $\gamma=17.0667$, the values of the ratio $Q/M$ 
are very close to unity. The largest values of $Q/M$ for case $1$ and case $2$ 
are respectively $0.999793$ and $0.999814$. This fact suggests that the 
quasiblack hole regime is about to be reached in both cases, and we 
investigate this point in detail later on.

\subsection{The structure dependence of the relativistic charged spheres on
the charge fraction}

\subsubsection{The mass of the spheres as a function of the charge fraction}

\begin{figure}[ht]
\centering
\includegraphics[scale=0.29]{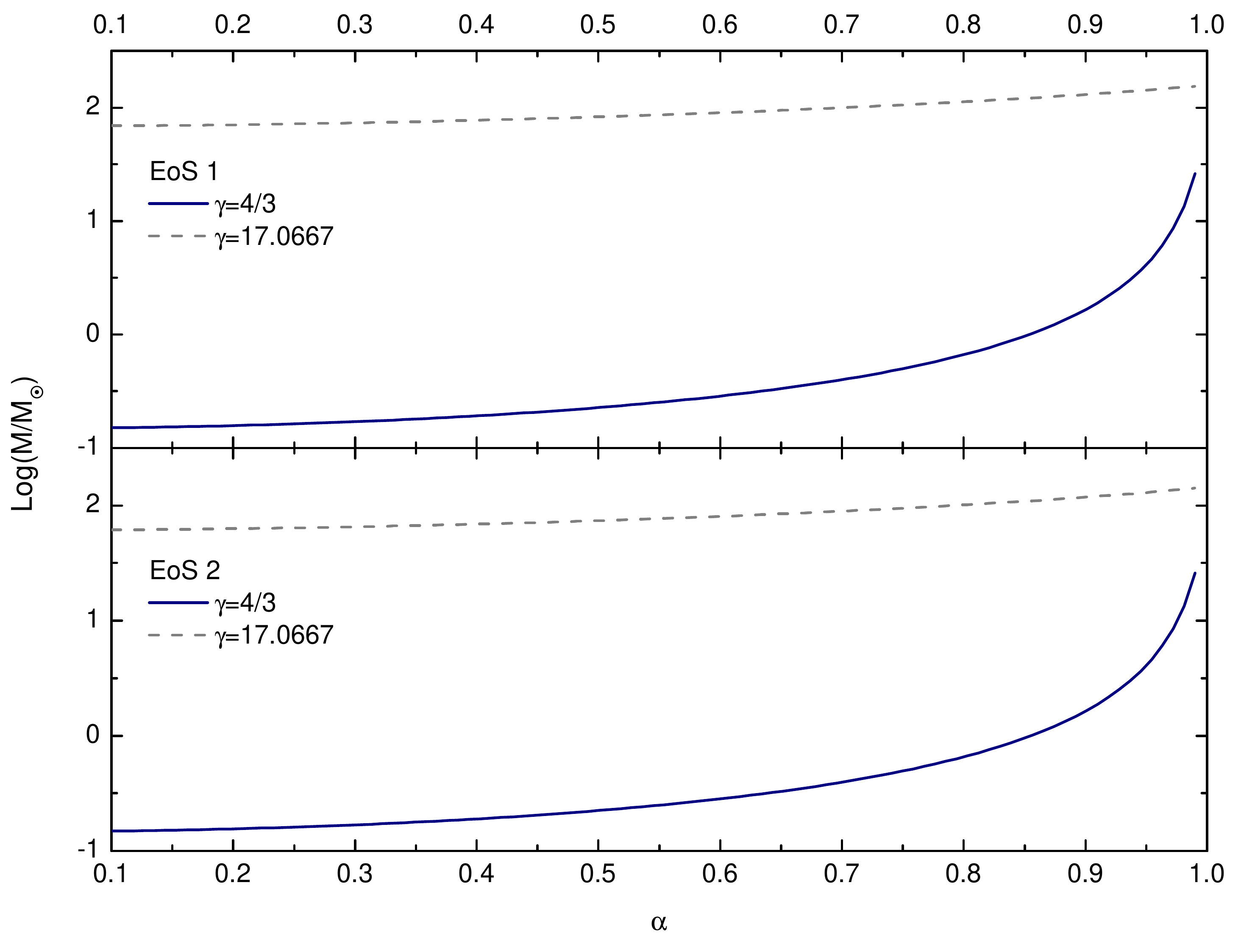}
\caption{Mass of the spheres versus the charge fraction $\protect\alpha$ for
two values of the polytropic exponent, $\protect\gamma= 4/3$ and 
$\protect\gamma=17.0667$. The top 
panel is for the polytropic equation of state (EoS $1$), with the
central energy
density $\protect\rho_c=1.78266\times10^{16}[\mathrm{kg/m^{3}}]$. The
bottom panel is for the relativistic polytropic equation of state (EoS
$2$), with the central rest mass density
$\protect\delta_c=1.78266\times10^{16}
[\mathrm{kg/m^{3}}]$.}
\label{alpha_M_Msol_EOSI_II}
\end{figure}

The ratio $M/M_{\odot}$ as a function of the charge fraction $\alpha$ is 
presented in Fig. \ref{alpha_M_Msol_EOSI_II}, for the case 1 with $ 
\rho_c=1.78266\times10^{16}[\mathrm{kg/m^{3}}]$ exhibited in the 
top panel, and for the case 2 with 
$\delta_c=1.78266\times10^{16}[\mathrm{kg/m^{3}}]$ displayed in the bottom 
panel. Two values of the polytropic exponent are considered,
$4/3$ and $17.0667$. In both fluid types, 
we see the smooth growth of the mass with the charge fraction
and also with the polytropic exponent.  The behavior of the curves 
indicate that the relation $M/M_\odot$ 
is essentially the same for the two equations of state. 

\subsubsection{The radius of the spheres as a function of the charge
fraction}

\begin{figure}[ht]
\centering
\includegraphics[scale=0.31]{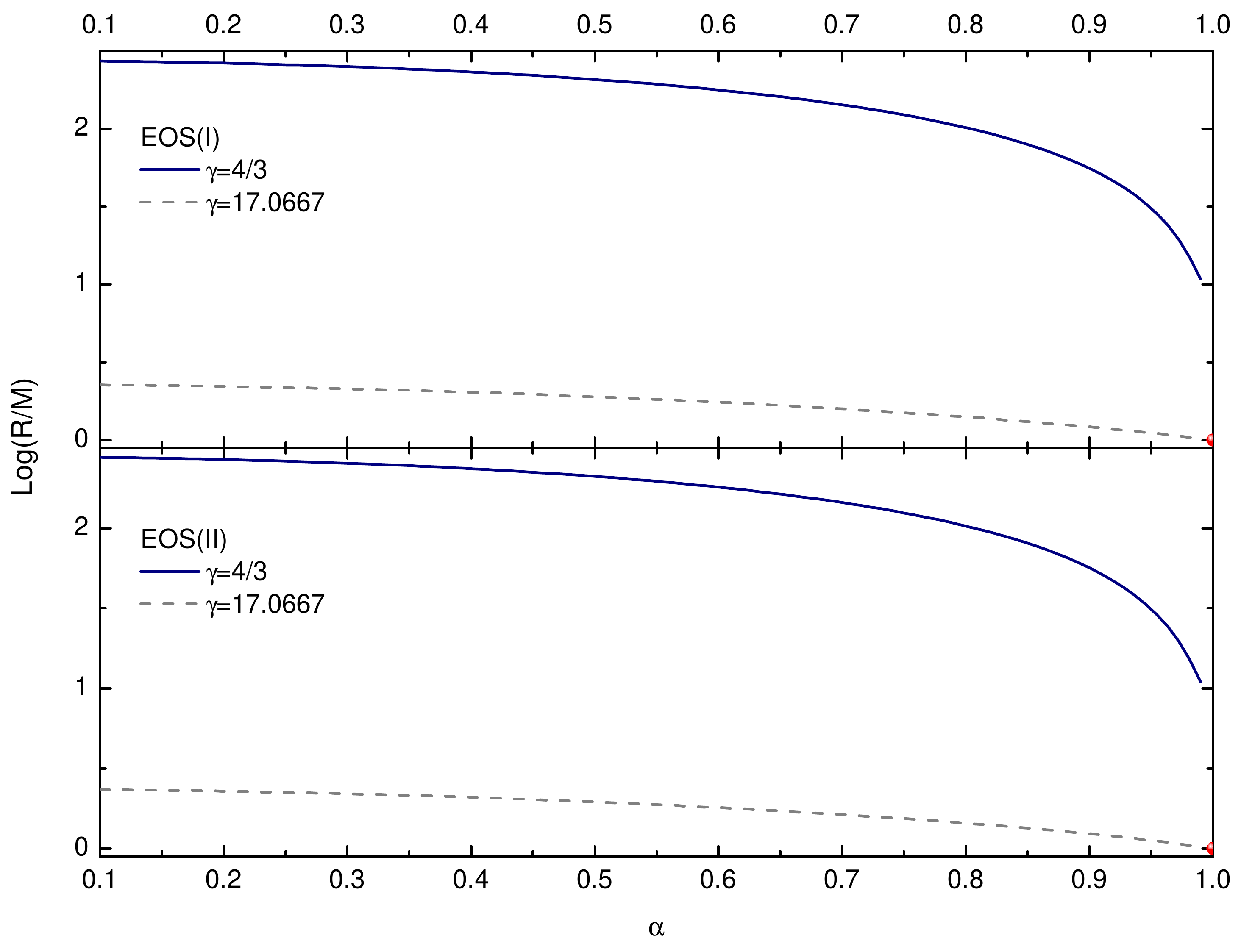}
\caption{The ratio $R/M$ as a function of the charge fraction for 
two values of the polytropic exponent, $\gamma=4/3$ and $\gamma=17.0667$, as 
indicated. The results for the  polytropic 
equation of state (EoS $1$) are shown in the top panel, while the results for 
the relativistic polytropic equation of state (EoS $2$) are shown in the 
bottom panel.}
\label{alpha_R_M_EOSI_II}
\end{figure}

 The radius to mass ratio ($R/M$) as a function of the charge fraction 
$\alpha$ for two values of the polytropic exponent, $\gamma= 4/3$ and 
$\gamma=17.0667$, is presented in Fig.~\ref{alpha_R_M_EOSI_II}. The top 
panel is for the EoS $1$ \ref{EoS1} with the central density
$\rho_c=1.78266\times10^{16}[\mathrm{kg/m^{3}}]$, while the bottom panel  
is for the EoS $2$ \ref{EoS2} with the central rest mass 
density $\delta_c=1.78266\times10^{16}[\mathrm{kg/m^{3}}]$. It is seen from 
the figure that, for both equations of state, the ratio $R/M$ 
decreases with the increment of the charge fraction $\alpha$. Note also that 
in the extreme case, with $\alpha=0.99$ and  $\gamma=17.0667$, the ratio 
$R/M$ is  close to unity for both equations of state. In fact, the values of 
$R/M$ are $1.02676$ and $1.02514$ in the EoS $1$ case  and in the EoS $2$ 
case, respectively.

\subsubsection{The charge of the spheres as a function of the charge
fraction}

\begin{figure}[ht]
\centering
\includegraphics[scale=0.29]{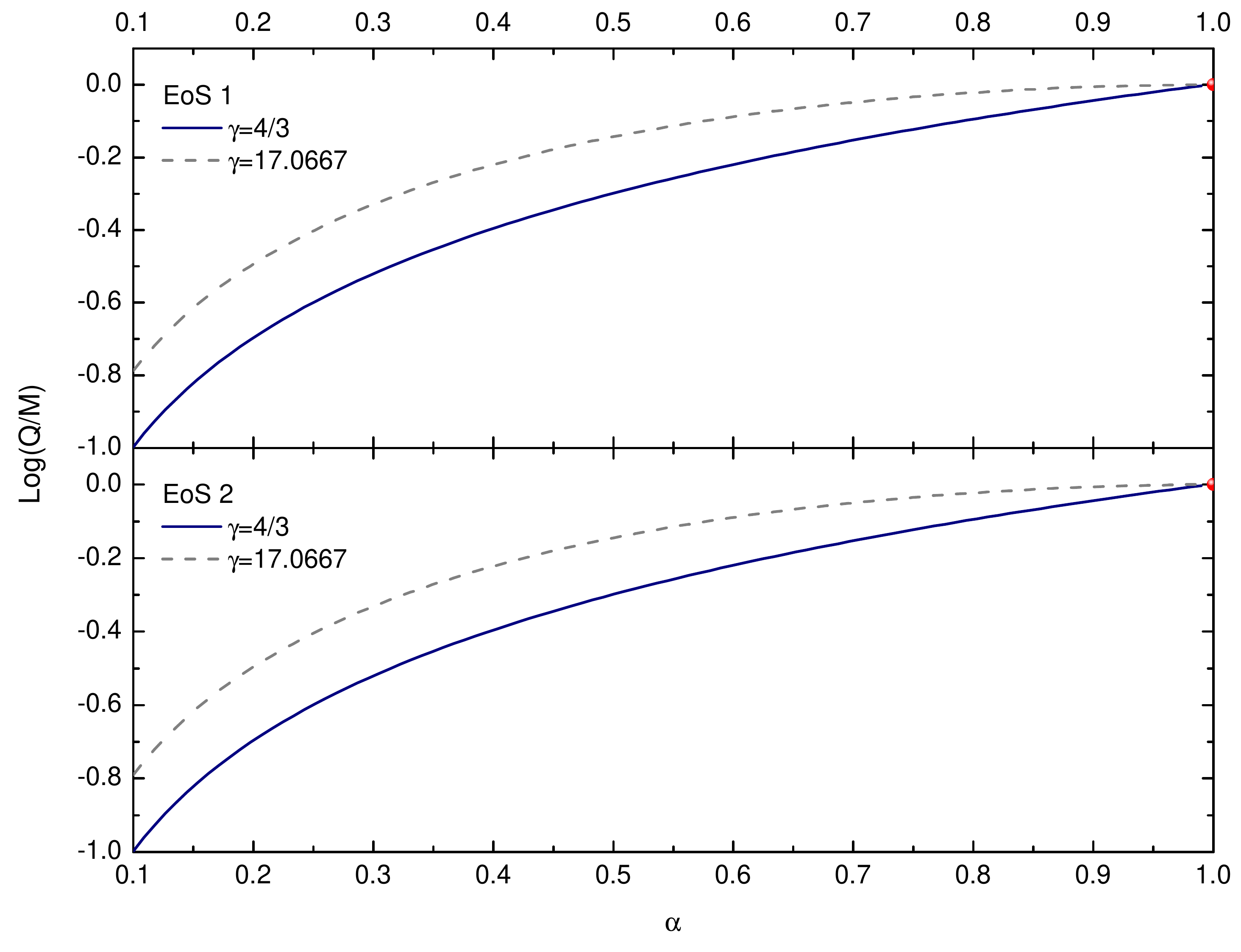}
\caption{The ratio $Q/M$ versus the charge fraction for $\gamma=4/3 
$ and $\gamma=17.0667$. The top panel is drawn in the case of the polytropic 
equation of state (EoS $1$), and the bottom panel in the case of the 
relativistic polytropic equation of state (EoS $2$).}
\label{alpha_Q_M_EOSI_II}
\end{figure}

The charge to mass ratio $Q/M$ versus the charge fraction $\alpha$
is presented in Fig.~\ref{alpha_Q_M_EOSI_II}, 
for two values of the polytropic exponent, $\gamma=4/3$ and $17.0667$. The 
top panel shows the results for the EoS $1$ with central energy density
$1.78266\times10^{16}[\mathrm{kg/m^{3}}]$. The bottom panel shows the results 
for EoS $2$ with the central rest-mass density
$1.78266\times10^{16}[\mathrm{kg/m^{3}}]$. The curves indicate that $Q/M\to 1$
for $\alpha \to 1$. As discussed below, this signals the facts that the 
Buchdahl-Andr\'easson bound and the quasiblack hole limit is about to be reached
for large charge fraction.

\section{Properties of relativistic polytropic fluid spheres with infinitely
large polytropic exponent}\label{sect-largegamma}

\subsection{Large polytropic exponent and incompressible fluids}

Analyzing equations EoS $1$ \eqref{EoS1} and EoS $2$ \eqref{EoS2}, 
it can be seen that in both cases a small $\gamma$ provides a low pressure
and a large $\gamma$ leads to a high pressure.

As shown above, at low fluid pressures (low exponents $\gamma$) the EoS $1$ 
is equivalent to the EoS $2$. On the other hand, at large fluid
pressures (large exponents $\gamma$), EoS $1$ and EoS $2$ yield
completely different results. In order to highlight the 
differences between these two equations of state for large values of
$\gamma$, a comparison between them in that region must be performed.
Even though the analysis of the polytropic equation of state (EoS $1$) in
such a limit was developed in \cite{ALZ}, for convenience we rewrite the
relevant parts of that analysis here. Let us call $p_{0}$ a
particularly chosen value of the central pressure and consider it as
a normalization factor for the pressure. Hence, using
Eq.~\eqref{EoS1} it follows,
\begin{equation}
\lim_{\gamma\rightarrow\infty}{\dfrac{p}{p_0}} =
\lim_{\gamma\rightarrow\infty }\left(\dfrac{\rho}{\rho_0}\right)^\gamma
              =\left\{\begin{array}{l}
		\infty, \quad {\rm if}\quad \rho > \rho_0;\\
		0, \quad {\rm if}\quad \rho < \rho_0;\\
              \end{array} \right. \label{plimit1}
\end{equation}
and
\begin{equation}
\lim_{\gamma\rightarrow\infty}{\dfrac{\rho}{\rho_0}} =
\lim_{\gamma\rightarrow\infty }\left(\dfrac{p}{p_0}\right)^{1/\gamma}
              = 1 . \label{rholimit1}
\end{equation}
where  $p_{0} = w \rho_{0}^\gamma$.\\
This limit conducts to an incompressible (constant energy-density $\rho$)
fluid, as in the Schwarzschild interior solution
\cite{incom_schwarzschild}, besides the addition of a constant electric
charge density, since we have also 

\begin{equation}
\lim_{\gamma\rightarrow\infty}{\dfrac{\rho_e}{\rho_{0e}}} =
\lim_{\gamma\rightarrow\infty}{\dfrac{\rho}{\rho_0}}=
\lim_{\gamma\rightarrow\infty }\left(\dfrac{
p}{p_0}\right)^{1/\gamma}               = 1 . \label{rhoelimit1}
\end{equation}
Such an electrified Schwarzschild interior solution was investigated in
\cite{ALZ2014}.

Now the fluid quantities in the case of EoS $2$  are normalized as
$p/p_{0}^{*}$, $\rho/\rho_{0}^{*}$ 
and $\delta/\delta_{0}^{*}$, where $\delta_{0}^{*}$, $\rho_{0}^{*}$ and
$p_{0}^{*}$ are normalization factors. These factors are related
by $p_{0}^{*}=\omega{\delta_{0}^{*}}^{\gamma}$ and
$\rho_{0}^{*}=\delta_{0}^{*}+p_{0}^{*}/(\gamma-1)$. 
Hence, we get
\begin{equation}
\lim_{\gamma\rightarrow\infty}{\dfrac{p}{p^{*}_{0}}} =
\lim_{\gamma\rightarrow\infty
}\left(\dfrac{\delta}{\delta^{*}_{0}}\right)^\gamma
              =\left\{\begin{array}{l}
        0, \quad\;\; {\rm if}\quad \delta < \delta^{*}_{0},\\
        \infty, \quad {\rm if}\quad \delta > \delta^{*}_{0},\\
              \end{array} \right. \label{plimit}
\end{equation}
and we have
\begin{equation}
\lim_{\gamma\rightarrow\infty}{\dfrac{\rho}{\rho^{*}_{0}}}=
\lim_{\gamma\rightarrow\infty}{\dfrac{\rho_e}{\rho^{*}_{e0}}}=
              \left\{\begin{array}{l}
         \dfrac{\delta}{\delta^{*}_{0}}, 
         \quad {\rm if}\quad \delta < \delta^{*}_{0},\\
        \infty, \quad {\rm if}\quad \delta > \delta^{*}_{0},\\
              \end{array} \right. \label{rholimit}
\end{equation}
with
\begin{equation}
{\dfrac{\rho}{\rho^{*}_{0}}} =
\dfrac{\left(\dfrac{\delta}{\delta^{*}_{0}}\right)}{1
+\dfrac{\bar\omega}{\gamma-1}} +
\dfrac{\left(\dfrac{\delta}{\delta^{*}_{0}}\right)^{\gamma}} 
{1 +\dfrac{\gamma-1}{\bar\omega}}
              , \label{rhonorm}
\end{equation}
where $\bar\omega = 1.47518\times 10^{-3}$ and we used Eq.~\eqref{omega}.

For the normalized rest mass density we get
\begin{equation}
\lim_{\gamma\rightarrow\infty}{\frac{\delta}{\delta^{*}_{0}}} =
\lim_{\gamma\rightarrow\infty
}\left(\frac{p}{p^{*}_{0}}\right)^{1/\gamma}
              = 1 . \label{deltalimit}
\end{equation}

Therefore, the limit of high polytropic exponents of the relativistic 
equation of
state (EoS 2) does not yield an incompressible fluid. It gives a constant
rest-mass density, and in the instance when the pressure may assume
arbitrarily large values, it gives an infinitely large energy
density too, and it gives a constant energy density in a second instance
when the pressure vanishes. This second situation is not interesting
for the present analysis. 

It is also worth mentioning that, since we assume the relation
$\rho_e=\alpha\rho$, the conditions given by Eq.~\eqref{rholimit} is also
fulfilled by the charge density in its normalized form, 
$\rho_e/\rho^{*}_{e0}$. 
Notice also that the polytropic constant $\omega$ plays an important role in
the normalization of the relativistic polytropic equation of state. Since
it depends upon $\gamma$, the normalization adopted according to 
Eq.~\eqref{omega} implies the results presented in Eqs.~\eqref{rholimit}
and \eqref{rhonorm}.

On the basis of results previously reported in this work, we know that when
$\rho_c<\rho_0$ and $\delta_c<\delta^{*}_{0}$ in the cases $1$ and $2$,
respectively, there are no equilibrium solutions for 
the polytropic spheres (see also \cite{ALZ}) and neither for the
relativistic polytropic spheres with infinitely large polytropic exponents.
On the other hand, as shown in \cite{ALZ} and confirmed in the present
study, in the limit $\gamma\longrightarrow\infty$  for case $1$, when
$\rho_c>\rho_0$, we have that the polytropic stars have constant energy
densities and infinitely large central pressures. For these 
(non-relativistic) polytropic star configurations it was found that the 
Buchdahl bound is saturated, thus, in the limit o zero electric charge, reaching 
the limit $R/M=9/4$. In turn, from the results presented
here for the EoS $2$ (case $2$), in the limit $\gamma\longrightarrow\infty$,
and with $\delta_c>\delta^{*}_{0}$, the relativistic polytropic star
configurations have both the central pressures and the central energy densities
becoming infinitely large. For these relativistic polytropic configurations, 
we have that the Buchdahl bound is far from being saturated.
\begin{figure}[h]
\centering
\includegraphics[scale=0.29]{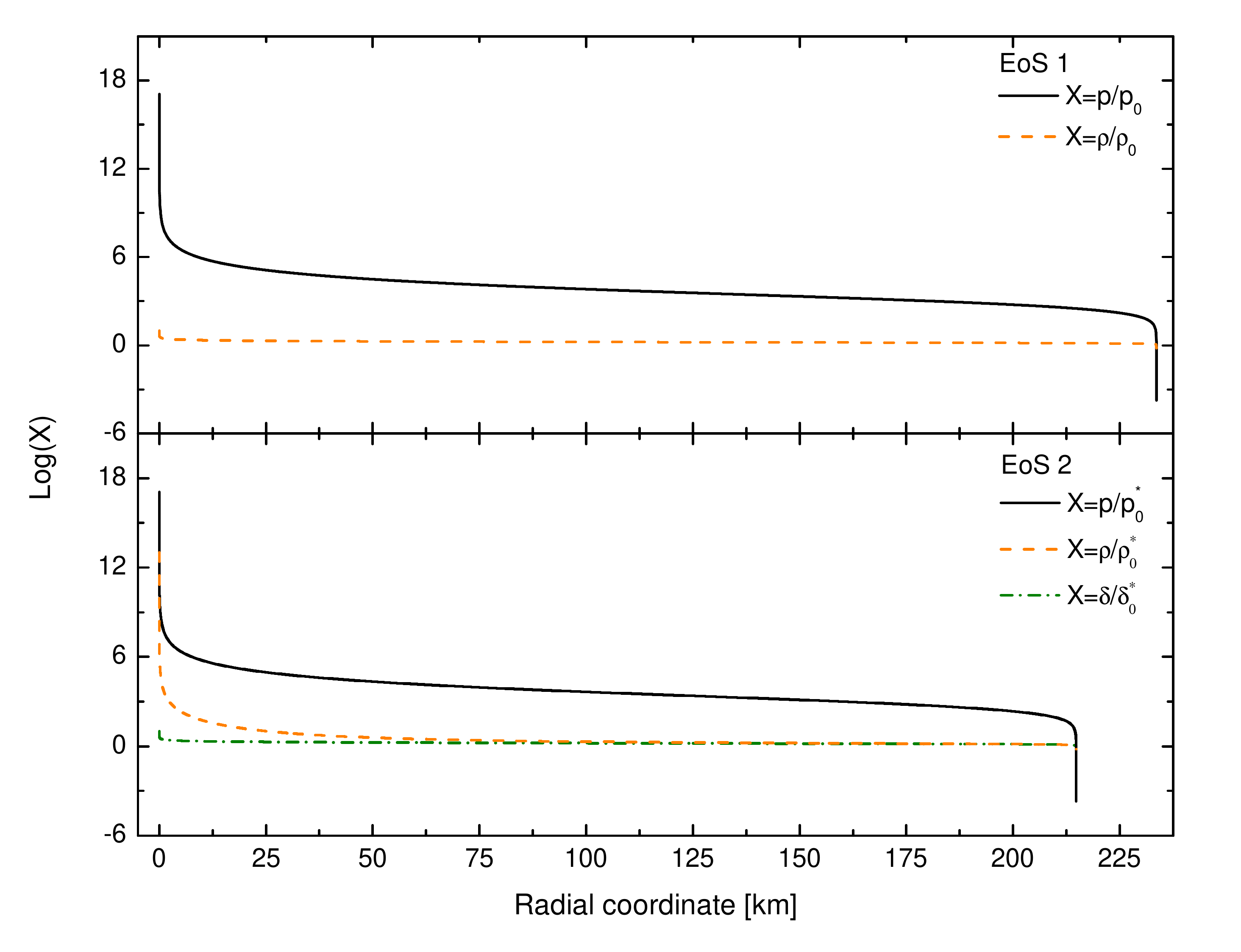}
\caption{Top panel: The radial dependency of the normalized functions 
$p(r)/p_0$ and $\rho(r)/\rho_0$ with
$\rho_c=1.78266\times10^{16}[\mathrm{kg/m^{3}}]$ 
in case $1$. Bottom panel: The radial dependency of the normalized
$p(r)/p_{0}^{*}$, $\rho(r)/\rho_{0}^{*}$ and 
$\delta(r)/\delta_{0}^{*}$ with
$\delta_c=1.78266\times10^{16}[\mathrm{kg/m^{3}}]$ in case $2$. In both
cases it was used $\alpha=0.99$ and $\gamma=17.0667$. The normalization
factors are
$\rho_0=\delta_{0}^{*}=1.78266\times10^{15}[\mathrm{kg/m^{3}}]$,
$p_0=p_{0}^{*}=2.62974\times10^{12}[\mathrm{kg/m^{3}}]$, and 
$\rho_{0}^{*}=1.78282\times10^{15}[\mathrm{kg/m^{3}}]$.}
\label{ra_coor_p_rho_delta}
\end{figure}
For the extremely charged case, however, we have seen that the use of the 
equations of state EoS $1$ and EoS $2$, with $\gamma\rightarrow\infty$, allows 
the stars to saturate the Buchdahl-Andr\'easson bound with $R/M\simeq1$. These 
solutions correspond to quasiblack holes. These important results are 
investigated in more detail in the following section.

The above analytical results regarding the equations of state 
of the non-relativistic polytropic, as well as of the  
relativistic polytropic fluid, at very large polytropic exponents can
be confirmed resorting to the numerical calculation.  For instance,
the behavior of the energy density $\rho(r)/\rho_{0}$ and of the fluid 
pressure $p(r)/p_{0}$ with the radial coordinate $r$ in case $1$ is shown in 
the top panel of Fig. \ref{ra_coor_p_rho_delta}.  As before, the central 
energy density for
EoS $1$ is $\rho_c=10\rho_0= 1.78266\times10^{16}[\mathrm{kg/m^{3}}]$. The
bottom panel of the figure shows the comportment of the energy density
$\rho(r)/\rho_{0}^{*}$, pressure $p(r)/p_{0}^{*}$,  and rest-mass density
$\delta(r)/\delta_{0}^{*}$ of the relativistic polytropic fluid (case $2$)
against the radial coordinate, with the central rest-mass density given by
$\delta_c=10\delta_0^*=1.78266\times10^{16}[\mathrm{kg/m^{3}}]$.
For every plots in Fig.~\ref{ra_coor_p_rho_delta}, the same charge fraction
$\alpha=0.99$ and the same polytropic exponent $\gamma=17.0667$ were used.

In both models the pressure inside the spheres starts with the same value at 
the center of the sphere and decreases monotonically with the radial 
coordinate. The pressure starts with a very high value at the origin, $r=0$,  
and reaches its minimum value on the surface of the sphere, at
$r=R$. On the other hand, note that the energy density for case $2$
has a completely different behavior when compared to case $1$. In case $1$,
the energy density is nearly constant, starting with 
$\rho(r)/\rho_0=10$ at $r=0$ and decreasing very slowly with $r$ until
the surface of the sphere at $r=R$, where it reaches its minimum value. For
case $2$, the energy density starts with the high value 
$\rho(r)/\rho^{*}_{0}=10^{18.3}$ at the center of 
the object, and varies rapidly with the radial coordinate to reach a 
value close to zero at the surface the object $r=R$.
Finally, in reference to the rest-mass density function, shown in the bottom 
panel of Fig. \ref{ra_coor_p_rho_delta}, we
see that it is approximately a constant, starting with the value
$\delta(r)/\delta^{*}_{0}=10$ at $r=0$ and decaying very slowly toward the
surface of the sphere $r=R$.

\subsection{Large polytropic exponents: The Buchdahl bound, the
Buchdahl-Andr\'easson bound, and the quasiblack hole limit}

The existence of upper bounds for compact objects is one of the remarkable 
predictions of general relativity.  The upper limit established by Buchdahl 
\cite{buchdahl} in the case of uncharged fluid spheres was extended to include 
electric charged fluid spheres by Andr\'easson \cite{andreasson_charged}. Our 
main concern here is testing these bounds for the relativistic polytropic 
spheres (case 2). So, we search in the parameter space, namely varying the 
central mass-density $\delta_c$, the charge fraction $\alpha$, and the 
polytropic exponent $\gamma$, for the most compressed objects. The outcome of 
such a search is that the extremely compressed spheres are found for large 
polytropic exponents. 
The central mass-density is not important, while the 
charge fraction is relevant but not essential since the Buchdahl bound is 
found for zero charge. The extremely compressed objects are found for large 
$\gamma$, but the compactness ratio $R/M$ depends also on $\alpha$, 
varying from $R/M = 9/4$ at $\alpha =0$ to $R/M\simeq 1$ for $\alpha =0.99$.

\begin{figure}[ht]
\centering
\includegraphics[scale=0.29]{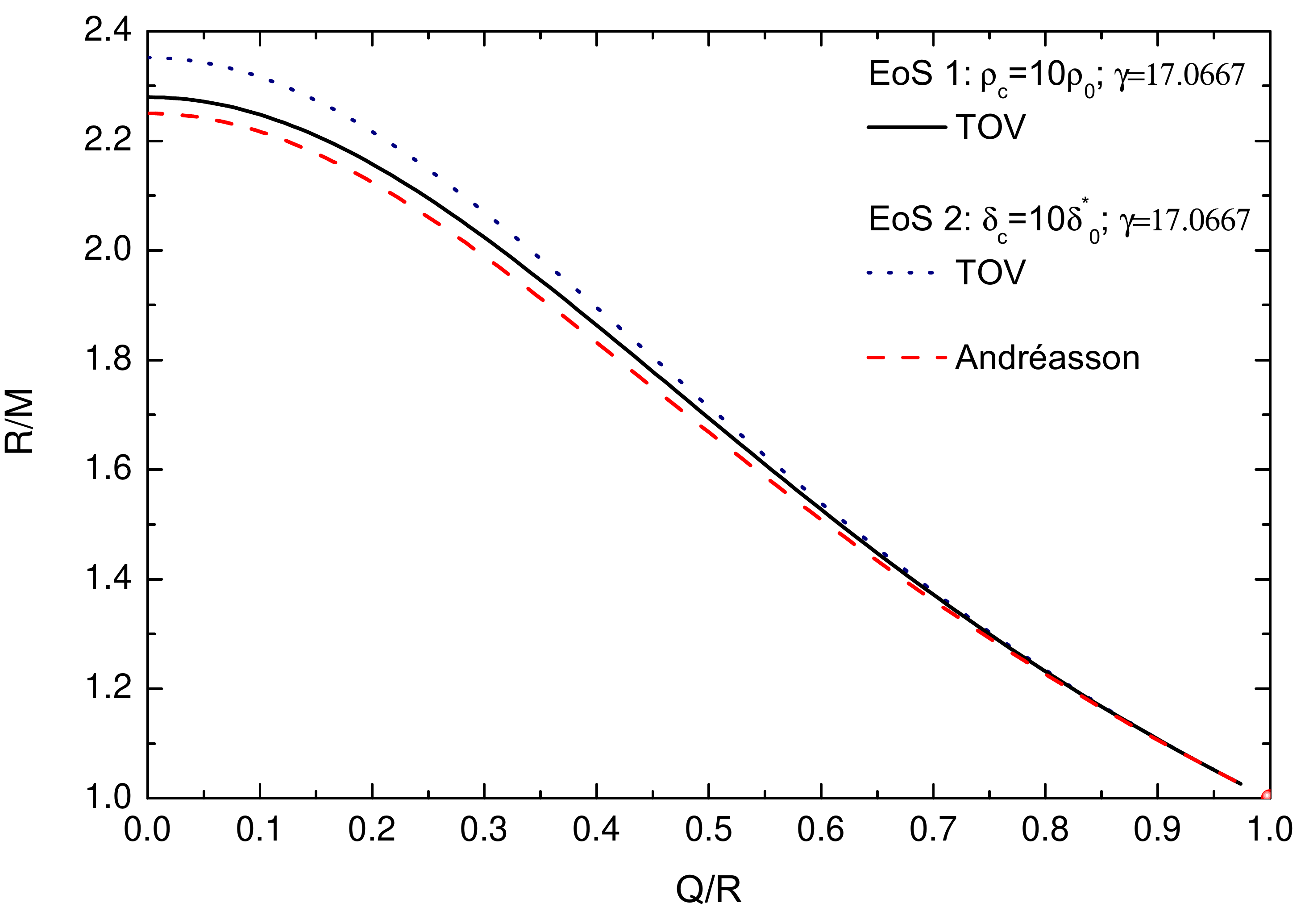}
\caption{The most compressed objects found numerically for EoS $1$ with
$\rho_c=1.78266\times10^{16}[\mathrm{kg/m^{3}}]$, and for EoS $2$ with
$\delta_c=1.78266\times10^{16}[\mathrm{kg/m^{3}}]$, as indicated.
The Buchdahl-Andr\'easson bound is also shown for comparison
(dashed line). This bound is saturated only in the limit of large charge
faction, $\alpha\rightarrow 1$, for which the quasiblack hole limit is
reached. }
\label{fig_BA-bound}
\end{figure}

Figure~\ref{fig_BA-bound} shows the behavior of the ratio $R/M$ as a 
function of $Q/M$ for the most compressed stellar static objects that follow 
from the EoS $1$ (solid line) with $\rho_c=1.78266\times10^{16}[\rm kg/m^3]$
and $p_c=2.62974\times10^{12}[\rm kg/m^3]$, and from the EoS $2$ (dotted line)
with $\delta_c=1.78266\times10^{16}[\rm kg/m^3]$ and 
$p_c=2.62974\times10^{12}[\rm kg/m^3]$. The two curves are drawn for the 
highest value of the polytropic exponent that yielded trusted numerical 
results, $\gamma = 17.0667$.

For the sake of comparison, the upper limit of the theoretical prevision by 
Andr\'easson (dashed line), which is given by the relation 
\begin{equation} \label{andrea-bound}
 \frac{R}{M}\geq\frac{9}{\left(1+\sqrt{1+3Q^2/R^2}\right)^{2}}\, ,
\end{equation}
is also depicted in Fig.~\ref{fig_BA-bound}.
This equation reproduces the Buchdahl bound for $Q=0$, viz, $R/M \geq 9/4$, 
and deliver the extremal compactness for $Q=M$, i.e., $2.25\geq R/M \geq 1$.
Notice that the two curves for the polytropic spheres (the solid and the 
dotted curves) appear above the Buchdahl-Andr\'easson bound for all $Q/R$.

In the limit of zero charge, 
$Q/R\rightarrow 0$, the ratio $R/M$ for the non-relativistic polytropic 
spheres (case 1) approaches the upper limit of the Buchdahl bound $R/M \to 
9/4=2.25$ better than for the relativistic polytropic spheres (case 2).
Noticing that the limit of infinitely high polytropic exponents yields an 
incompressible fluid in case $1$, 
the previous results of Ref.~\cite{ALZ2014} (see also \cite{ALZ}) assure that 
the Buchdahl bound is saturated by the uncharged fluid spheres in such a case.
The numerical 
calculation does not reach the ceiling value $M/R=9/4$ since the method 
employed here does not allow to go beyond $\gamma=17.0667$. In the same limit 
of zero electric charge, the curve for the relativistic polytropic spheres
(case 2) fails to converge to $R/M = 9/4$. In fact, the values shown in 
Fig.~\ref{fig_BA-bound} at $Q=0$ are $R/M\simeq 2.28$ for the 
EoS $1$, and $R/M\simeq 2.35$ for the EoS $2$.  Thus, the Buchdahl 
bound is not saturated by the uncharged fluid spheres with the 
relativistic polytropic equation of state (case $2$).

On the other side of the parameter space, for 
large charge fractions, $\alpha\to 1$, the two curves for charged spheres 
converge to the Buchdahl-Andr\'easson line. This means that the two 
equations of state model very compressed objects that saturate the 
Buchdahl-Andr\'easson bound in such a limit. The three lines converge to the 
quasiblack hole limit $R=M=Q$. As a matter of fact, in the cases 
analyzed here, the maximum value of the charge faction is $\alpha=0.99$ 
rather than $\alpha=1.0$, since we have not found static equilibrium 
solutions (the numerical method does not converge) for $\alpha$ larger than 
$0.99$. For this value of $\alpha$ we have found $R/M\simeq 
Q/M\simeq1.02676$ in case $1$, and $R/M\simeq Q/M\simeq1.02514$ in
case $2$.

Let us stress that in Fig.~\ref{fig_BA-bound} the three lines showed are very 
close to each other in the region $Q/R\simeq 1.0$. However, these lines do not 
coincide, thus, indicating that the values of $R/M$ shown by the dotted line 
and the solid line are near but always larger than those shown by the dashed 
line. The numerical results indicate that the three lines shall coincide just 
in the limit $\alpha\rightarrow 1$ with $\gamma\rightarrow\infty$.

\section{The speed of sound in relativistic polytropic 
charged spheres}\label{sound-section}

The aim here is to verify the limit, if there is one, where causality may
be violated, as done in Ref.~\cite{ALZ} for the non-relativistic polytropes.

The speed of sound in a compressible fluid is determined through 
the relation $c_s^{2}=dp/d\rho$. For the non-relativistic polytropic equation 
of state (EoS $1$), this gives 
\begin{equation} \label{sound-case1}
 {c}_s^2=\gamma\, \omega\, \rho^{\gamma-1}= \dfrac{\gamma p}{\rho}.
\end{equation}
For the relativistic
polytropic equation of state (EoS $2$), we get 
\begin{equation}\label{sound-case2}
{c}_s^{2}=\frac{dp}{d\rho}=\frac{\gamma p}{\rho+p}.
\end{equation}

First we comment on the dependence of the speed of sound in terms of the 
energy density. It is well known that, for any $\gamma>1$, EoS $1$ violates 
the causality condition ($c_s^2\leq 1$) for large energy densities. Namely, if 
$\gamma\,\omega\,\rho^{\gamma-1} > 1$ then,  
since $\omega$ is a positive constant parameter,  for 
sufficiently large $\rho$, it gives $c_s^2>1$ for any given $\gamma>1$ and 
$\omega>0$. On the other hand, it is also known that EoS $2$ does not violate 
the constraint $c_s^2\leq 1$ for large $\rho$. In fact, by taking the limit of 
large energy densities of the ratio $\gamma p/(\rho+p)$ it yields $c_s^2 = 
\gamma\left(\gamma-1\right)/2$. One then sees that $c_s^2 $
equals unity for $\gamma =2$. Therefore, 
as also known, the relativistic equation of state does not violate causality 
for $\gamma$ in the interval $1 \leq \gamma \leq 2$.

\begin{figure}[h]
\centering
\includegraphics[scale=0.29]{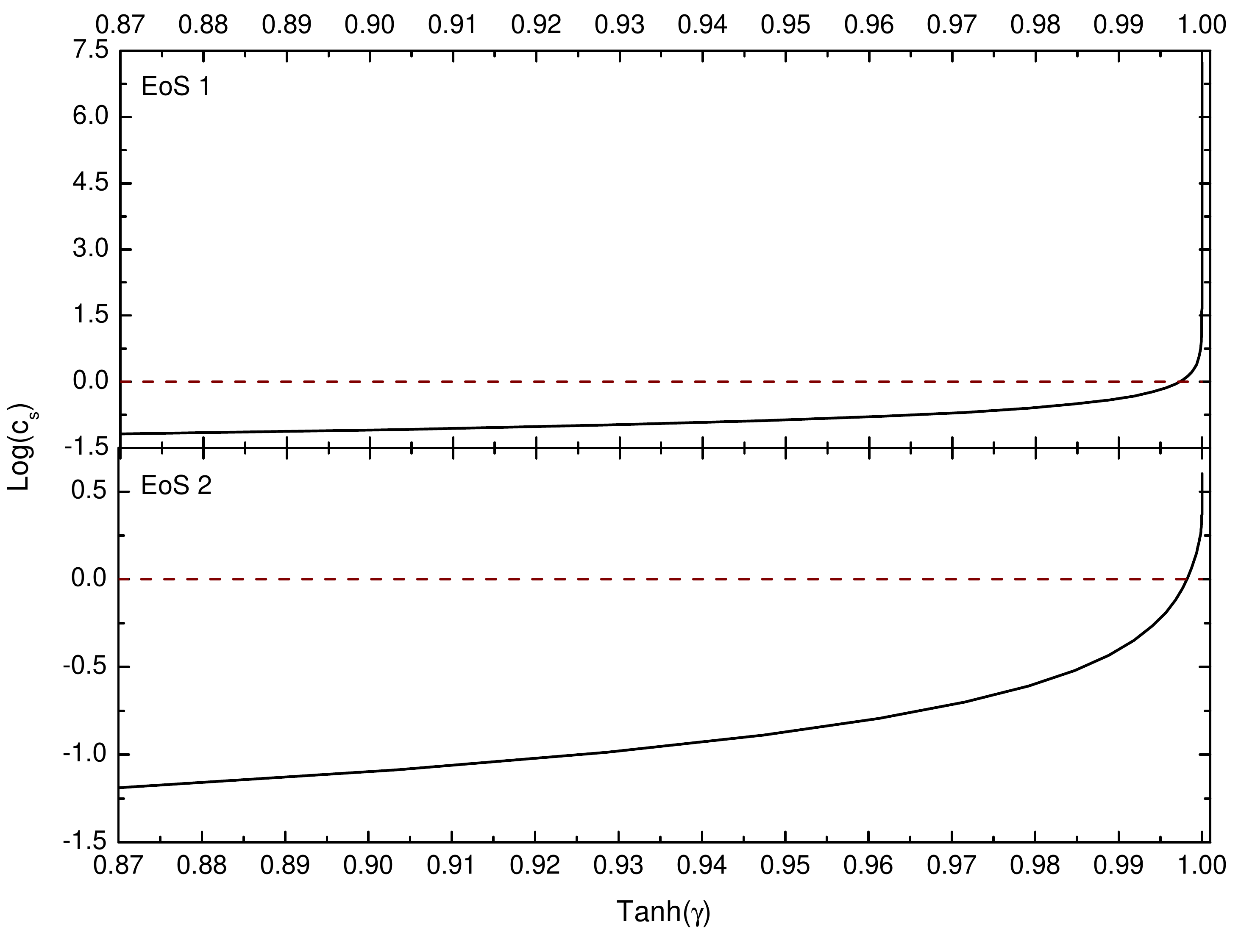}
\caption{The central speed of sound versus the polytropic exponent for the 
EoS $1$ with $\rho_c=1.78266\times10^{16}[{\rm kg/m^{3}}]$ (top panel), and 
for EoS $2$ with $\delta_c=1.78266\times10^{16}[{\rm kg/m^{3}}]$ (bottom 
panel).}
\label{Tanh_gam_log_c_s_EOSI_II}
\end{figure}

Now we comment on the dependence of the speed of sound in terms of the
polytropic exponent. Since the speed of sound decreases toward the surface of 
the sphere, as happens to the pressure, to see if the velocity of sound 
exceeds the speed of light it is only necessary to analyze the speed of sound 
in the center of the objects. The dependence of the central 
(at $r\to 0$) speed of sound upon the polytropic exponent is shown in 
Fig.~\ref{Tanh_gam_log_c_s_EOSI_II}, where we plot the results for the 
non-relativistic polytropic equation of state (EoS 1)) with  $\rho_c=1.78266\times 
10^{16}[{\rm kg/m^{3}}]$ (top panel), and for the relativistic polytropic 
equation of state (EoS 2) with $\delta_c=1.78266\times10^{16}[{\rm kg/m^{3}}]$ (bottom 
panel). We determine that the speed of sound $c_{s}$ in the center of the 
spheres reaches the speed of light at $\gamma\simeq 3.31120$ for the EoS $1$, and 
at $\gamma\simeq 3.52364$ for the EoS $2$.

Despite the fact that, in both kinds of fluids, the sound speed surpasses the speed 
of light for sufficiently high values of the Polytropic exponent $\gamma$, 
these solutions are interesting because, in such a limit, the fluids
become incompressible and the quasiblack hole limit is reached.

\section{The quasiblack hole limit of a relativistic
polytropic charged sphere}\label{qbh-section}

\subsection{Basic properties and the quasiblack hole limit}

From the results reported in Ref.~\cite{ALZ} and reproduced here in the top
panels of Figs.~\ref{Tanh_gam_R_M_EOSI_II} and \ref{Tanh_gam_Q_M_EOSI_II},
it is verified that the non-relativistic polytropes (EoS $1$) with charge fraction 
$\alpha=0.99$ and polytropic exponent $\gamma=17.0667$ are very close to the 
quasiblack hole configuration. Here, we have verified that a similar situation 
happens for the relativistic charged polytropes (EoS $2$), as seen in the 
bottom panels of the cited figures. For $\alpha=0.99$ and $\gamma=17.0667$ one 
has $R\simeq M\simeq Q$ (with $R$, $M$ and $Q$ expressed in geometric units), 
indicating that these objects are also quasiblack holes. In fact, it was 
argued in \cite{ALZ} that charged polytropic spheres in the limit of 
infinitely large polytropic exponent and charge fraction reaching unity are 
quasiblack holes. Now we check if the relativistic polytropic equation with 
$\alpha\rightarrow 1$ and $\gamma\longrightarrow\infty$ yields quasiblack 
holes too. For this purpose, following the defining properties of a static 
quasiblack hole put forward in Ref.~\cite{lz1}, the potential metrics $A(r)$ 
and $B(r)$ are analyzed.
In addition, we make a comparison of the behavior of the metric potentials
for both cases (the EoS $1$ and the EoS $2$), at the largest values $\alpha=
0.99$ and $\gamma = 17.0667$, as a function of the radial coordinate.
The central energy density and central
rest-mass density are $\rho_c = 10 \rho_0 = 1.78266\times10^{16}[{\rm
kg/m^{3}}]$ and $\delta_c = \delta_0^* =1.78266\times10^{16}[{\rm
kg/m^{3}}]$ for Eos $1$ and EoS $2$, respectively.

\begin{figure}[h]
      \centering
      \includegraphics[scale=0.29]{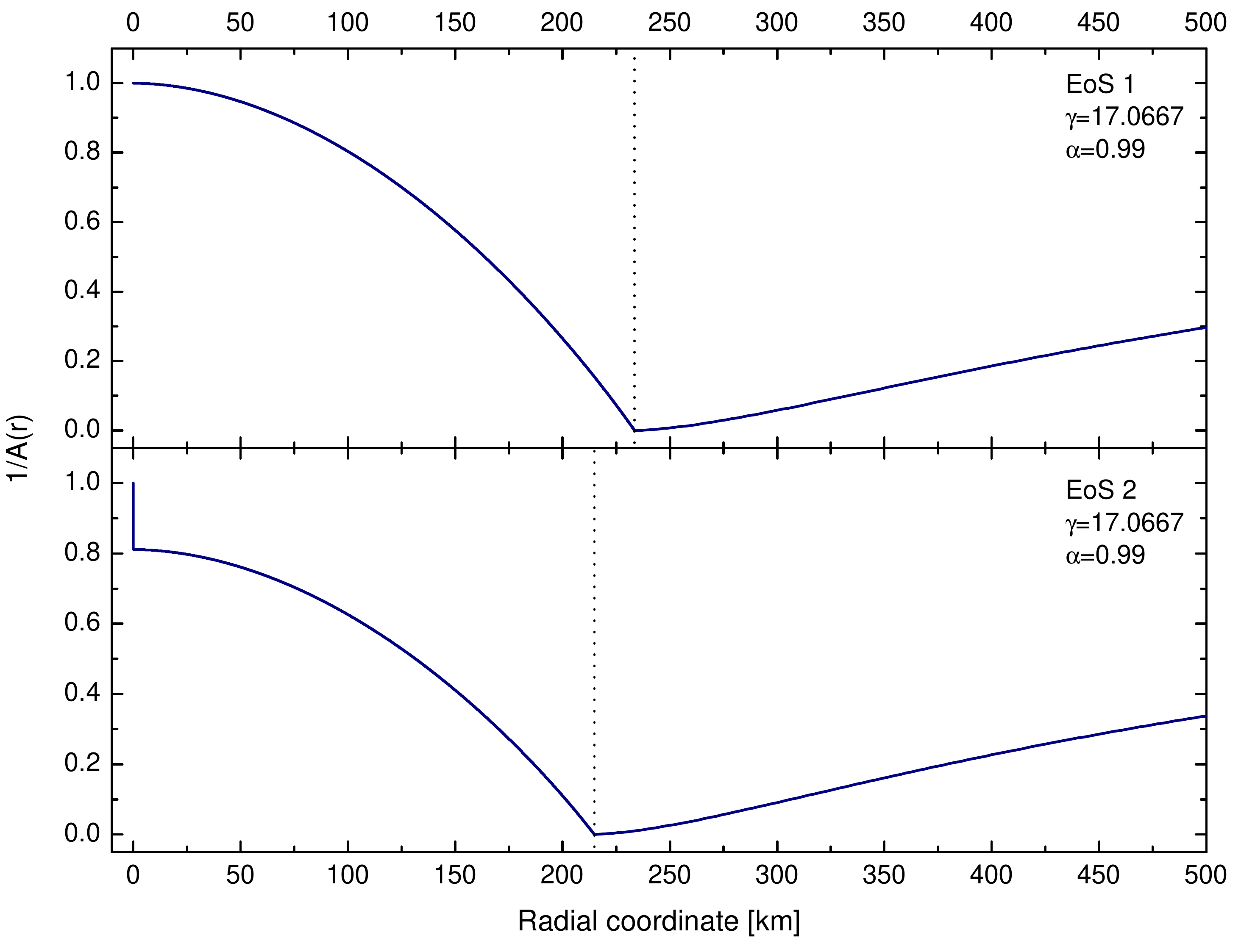}
      \caption{The metric function $A^{-1}(r)$ against the radial coordinate
  for the EoS $1$ with $\rho_c = 10\rho_0$ (top panel), and for the EoS $2$
with $\delta_c =  10 \delta_0^*$ (bottom panel), and with $\gamma=17.0667$
and $\alpha=0.99$ in both cases. The dotted vertical lines indicate the
surface of the spheres.}
      \label{radial_coor_vs_1_grr_099_170667}
\end{figure}

The inverse of the metric function $A(r)$ versus the
radial coordinate $r$ is plotted in 
Fig.~\ref{radial_coor_vs_1_grr_099_170667} for the EoS $1$ (top panel) and
the EoS $2$ (bottom panel). Near the origin ($ r\sim 0 $) 
there is a sharp difference between the two shown curves. 
In the case $2$, function $1/A(r) $ presents a jump from $1$ to about
$0.823$ at the central region. This is due to the fact that,
close to the center, the electric charge $q(r)$ and mass $m(r)$
grow rapidly with $r$ due to the very large values of the central energy and
charge densities. Nevertheless, in case $1$ their growth is smooth since the
respective densities are not very high. Function $A^{-1}(r)$ decreases with
the increasing of the radial coordinate,
reaching its minimum value, namely, $A^{-1}(R) \sim 0$,  at the
surface of the object. Such a small value signals that the object is close
to a quasiblack hole configuration. The interior metric is matched to the
exterior Reissner-Nordstr\"om metric, i.e., $A^{-1}(R)= 1- 2M/R + Q^2/R^2$, 
from what follows that the quasi-horizon is present.

\begin{figure}[h]
      \centering
      \includegraphics[scale=0.29]{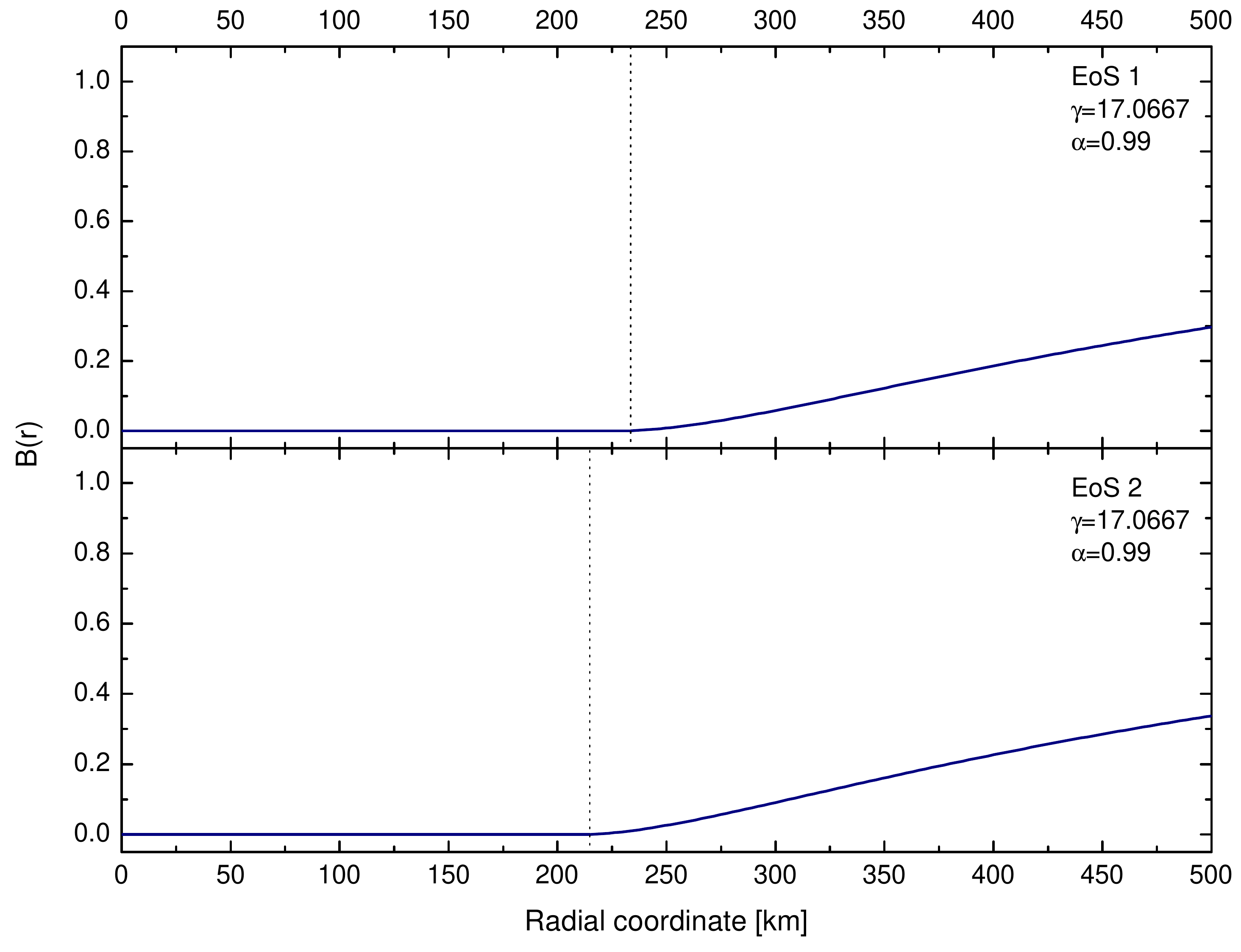}
      \caption{The metric function $B(r)$ against the radial coordinate
  for the EoS $1$ with $\rho_c = 10\rho_0$ (top panel), and for the EoS $2$
with $\delta_c = 10\delta_0^*$ (bottom panel), and with $\gamma=17.0667$
and $\alpha=0.99$
in both cases.}
      \label{radial_coor_vs_gtt_099_170667}
\end{figure}

The metric function $B(r)$ is shown in 
Fig.~\ref{radial_coor_vs_gtt_099_170667}, 
for the cases $1$ (top panel) and $2$ (bottom panel).  Note 
in the figure that function $B(r)$ assumes values close to zero 
in the interior of the sphere, i.e., 
$B(r)\rightarrow0$ in the whole interval
$0\leq r\leq R$. This feature also reveals that we are close to the
quasiblack hole configuration. Since the interior solution is 
matched to the exterior vacuum Reissner-Nordstr\"{o}m
solution, it follows we have $B(R)=A^{-1}(R)=1-2M/R+Q^2/R^2\sim 0$,
confirming once again the presence of a quasi-horizon.

Besides the defining properties of the metric potentials, as just checked,   
in the case of static charged spacetimes, another important property of
quasiblack holes is the existence of an extremal limit for the ratio $Q/M$.
As already mentioned, for large Polytropic exponent and charge fraction close 
to unity, we have found that the radius, the total mass, 
and the total charge of the relativistic polytropic spheres are close to each
other ($R\simeq M\simeq
Q$). Then, considering the two solutions of equation $F(r)=1-2M/r
+Q^2/r^2 = 0$, which are $r_{\pm}=M\pm\sqrt{M^{2}-Q^{2}}$, we get 
$r_+\simeq r_-\simeq M\simeq Q\simeq R$.
Moreover, the numerical analysis shows that the extremal bound $R=Q=M$ is
continuously approached with the increasing of the polytropic exponent and,
in particular, of the charge fraction $\alpha$. This means that the radius
of the charged matter distribution is reaching the gravitational radius from
above, $R\gtrsim r_{+}$, assuring the solution is regular, static, and very
close to extremality,
i.e., the quasiblack hole with pressure limit is being attained.

\begin{table*}
\begin{ruledtabular}
\begin{tabular}{ccccccccc}
$ $ & EoS & $\gamma$ & $M\times10^{5}\,[{\rm m}]$ & $Q\times10^{5}\,[{\rm m}]$ &
$R\times10^{5}\,[{\rm m}]$ & $R/M$ & $Q/M$\\\hline
A & $1$ & $17.0667$ & $2.27478$ & $2.27431$ & $2.33566$ & $1.02676$ & $0.999793$ \\
B & $2$ & $17.0667$ & $2.09502$ & $2.09463$ & $2.14769$ & $1.02514$ & $0.999813$\\
C & $2$ & $17.1109$ & $2.09662$ & $2.09623$ & $2.14929$ & $1.02512$ & $0.999814$
\end{tabular}
\caption{The values of the mass $M$, charge $Q$, and radius $R$ of the charged
polytropic spheres, in geometric units, with the corresponding 
values of $R/M$ and $Q/M$, for $\alpha=0.99$ and $\gamma=17.0667$
are shown in rows A and B, respectively, for case 1 and case 2. 
The values of $M$, $Q$, $R$, $R/M$, and $Q/M$, for the EoS $2$ (case 2) and for the 
polytropic exponent $\gamma=17.1109$ are shown in row C.}
\end{ruledtabular}
\end{table*}

Table~I presents the mass $M$, the charge $Q$, the radius $R$, and their
relations for the relativistic polytropes (EoS 2) with $\alpha=0.99$,  and
for two values of the polytropic exponent, $\gamma=17.0667$ and 
$\gamma=17.1109$. These are the highest values of $\gamma$
the numerical procedure yielded results
 without convergence problems.
For comparison, the same quantities for the EoS $1$ case
are also listed in the table (row A). By analyzing rows B and C it is seen that
$R/M$ and $Q/M$  are closer to unity in the EoS $2$ case than in the EoS $1$
case. Based on these results we note that the relativistic 
polytropic spheres with an infinitely large polytropic exponent and
charge fraction approaching unity attain the quasiblack 
hole limit, i.e., approach $R/M=Q/M=1.0$, faster than the
non-relativistic polytropic spheres do.

\subsection{The redshift at the surface of a quasiblack hole}\label{RS}

\begin{figure}[h]
\centering
\includegraphics[scale=0.29]{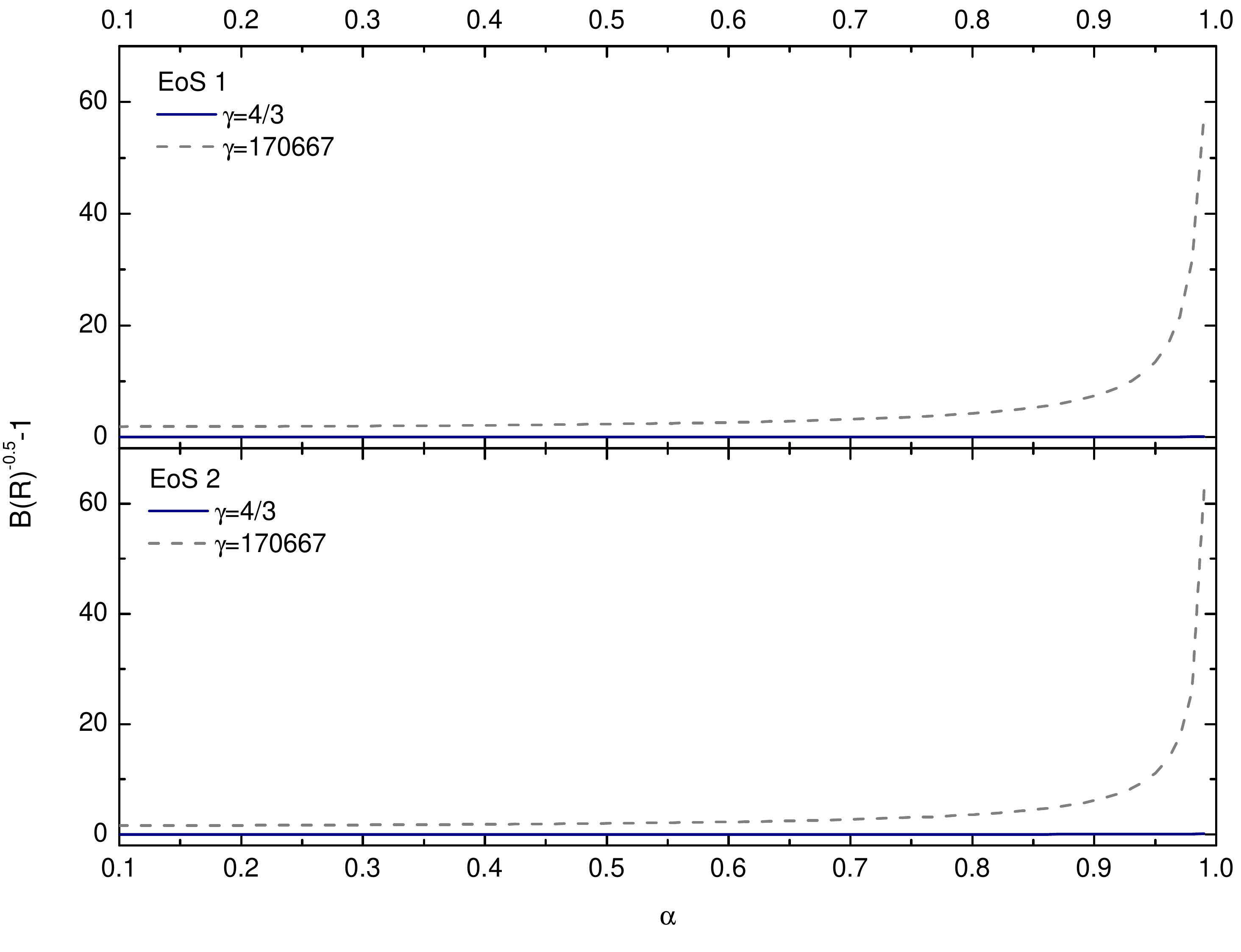}
\caption{The redshift function $B(R)^{-1/2}-1$ at the surface of the spheres 
for two values of the polytropic exponent, $\gamma=4/3$ 
and $17.0667$, for the EoS $1$ with $\rho_c=1.78266\times10^{16}[{\rm 
kg/m^{3}}]$ (top panel), and for the EoS $2$ with 
$\delta_c=1.78266\times10^{16}[{\rm kg/m^{3}}]$ (bottom panel). The charge 
fraction varies in the range $0<\alpha<1$.}
\label{alpha_redshift}
\end{figure}

To be complete, we calculate the quantity $B(R)^{-1/2}-1$ for both the 
non-relativistic and the relativistic charged polytropes, and taking
two values of the polytropic exponent, $\gamma=4/3$ and $17.0667$. 
The expression 
$B(R)^{-1/2}-1$ gives the redshift at the surface of the star, which is 
defined in the usual way by the fractional difference between the light wave 
frequency at the surface of the star (at $r=R$) with respect to infinity (at 
$r\to\infty$). The dependence of the redshift as a function of the charge 
fraction is plotted in Fig.~\ref{alpha_redshift} for the two equations of 
stated investigated in the present work. As expected from previous works on the 
non-relativistic polytropes \cite{lz1,ALZ2014}, the redshift at the surface of 
the quasiblack hole limit is infinitely large. Numerically we determine values 
of the redshift of about $100$ in the cases with $\alpha =0.99$ and 
$\gamma=17.0667$. Again the results for the EoS $2$ are very close to those for the 
EoS $1$ (see \cite{ALZ2014}), but the redshift is a little higher for the 
relativistic polytropes (EoS 2).

\section{Conclusions}\label{sec-conclusion} 

We compared the stellar structure configurations of charged objects made of a 
non-relativistic polytropic fluid [case 1, see Eq.~\eqref{EoS1}]
with those composed by a relativistic polytropic
fluid  [case 2, see Eq.~\eqref{EoS2}] in the Maxwell-Einstein theory. 
For the two cases analyzed, i.e., for the non-relativistic polytropic and 
relativistic polytropic cases, we used respectively the equation of state 
$p=\omega\rho^{\gamma}$ (EoS $1$, case $1$), and $p=\omega\delta^{\gamma}$, 
with $\delta=\rho-p/(\gamma-1)$ (EoS $2$, case $2$). The parameters $\omega$ 
and $\gamma$ represent respectively the polytropic constant and polytropic 
exponent. 
The chosen value of $\omega$ is such that for $\gamma=5/3$ the results found 
in \cite{raymalheirolemoszanchin,siffert,ALZ} for polytropic stars are 
reproduced.
The configurations studied are assumed to be composed by a spherically 
symmetric distribution of a charged perfect fluid, and by an exterior 
vacuum region described by Reissner-Nordstr\"om metric. 
We assumed a charge density profile directly proportional to the energy 
density, of the form $\rho_e=\alpha\,\rho$ (with $\alpha$ being the charge 
fraction).  
By varying the fundamental parameters of each model,
we analyzed some limits found in general relativity, such as the 
Chandrasekhar limit, the Oppenheimer-Volkoff limit, the Buchdahl bound, and the
Buchdahl-Andr\'easson bound and the quasiblack hole limit.

First the analysis was done by varying the central energy density
$\rho_c$ (for case 1), the central rest mass density $\delta_c$ (for case 2),  
and comparing the physical parameters (radius, mass, and charge) of
respective equilibrium solutions. A few different values of the polytropic 
exponent and of the charge fraction were considered in such an analysis. 
The study confirmed that the two equations of state yield significantly
different results just in the limit of high energy densities.

The configurations of the objects were also analyzed by varying the polytropic
exponent $\gamma$, from $4/3$ to a considerably high value.
In this situation, the central energy density (for case 1) and the central rest
mass density (for case 2) were kept fixed to ten times the normalization
values, i.e., $\rho_c=10\rho_0$ and $\delta_c=10\delta_{0}^{*}$,
with  $\rho_0=\delta_{0}^{*}= 1.78266\times10^{15}[\rm
kg/m^3]$. Using such values, we varied $\gamma$ from $4/3$ to 
$17.0667$ and $17.1109$, respectively, in case 1 and case 2. Higher values
of $\gamma$ introduced numerical convergence problems.
For the sake of comparison between the results in case $1$ and $2$, 
we considered $\gamma=17.0667$ as a maximum value for the polytropic 
exponent. For each one of the two equations of state, a detailed analysis of 
the equilibrium configurations was done, by calculating the radius, the mass, 
and the charge of each configuration, and then by comparing the results between 
the two cases. The results are very similar in both cases.
The main differences are related to the central energy density, which
goes with the pressure for high values of $\gamma$ in case 2, while it is
almost a constant in case 1.

The charge faction parameter $\alpha$ was varied from zero to very close to 
unity, $\alpha=0.99$. A value higher than this also  implied in numerical 
convergence problems. 
Again the structure of the resulting equilibrium solutions are almost the same
for both models of fluids.

In the regime of high polytropic exponents, we tested the various bounds for 
extremely compact objects. 
In fact, for the uncharged case ($\alpha=0.0$), in the case $1$, we have that for 
$\gamma=17.0667$ the Buchdahl bound is saturated, i.e., the Schwarzschild 
interior limit is attained. However, in case $2$ the Buchdahl bound is far from 
being saturated.  
In the extremely charged case ($\alpha=0.99$), and yet with $\gamma=17.0667$,
we have that the radius $R$, the mass $M$, and the charge $Q$ 
of the objects are approximately the same, $R\simeq M\simeq Q$. This result is
obtained for both equations of state.
This result together with the specific
characteristics of the potential metrics, i.e., $A^{-1}(R)\rightarrow0$ and 
$B(r)\rightarrow0$ with $r\leq R$, points the presence of a quasiblack hole.

The surface redshift of the extremely compact solutions, 
including the quasiblack hole limit, were analyzed. The results show higher
redshifts for relativistic polytropes (case 2) than for non-relativistic 
polytropes (case 1). 

The dependence of the sound speed $c_s$ on the polytropic exponent at the 
center of the compact objects was also studied. In both cases $c_s$ reaches 
values higher than the speed of light for sufficiently high polytropic indexes.

Finally, we emphasize that the aim of this work was to analyze the structure
of relativistic polytropes by comparing to non-relativistic polytropes, with 
particular interest in the upper bounds of compactness established within the
theory of general relativity. This is in complement of previous works by us
whose results were reported in Refs.~\cite{ALZ,ALZ2014}. The conclusion of this
investigation is that the Buchdahl-And\'easson bound is not saturated in full
neither by polytropic stars nor by incompressible stars. On the other hand, as
shown in Ref.~\cite{lemos_zanchin_Guilfoyle2015}, that bound is saturated by 
the Guilfoyle \cite{guilfoyle} solutions, which assumes different conditions
on the fluid quantities. This result suggests that a different equation of 
state for the charged fluid, associated to an alternative charge density 
profile, may lead to solutions that saturate that important bound, besides
reaching the quasiblack hole limit. The analysis of such situations 
is left for future investigations.

\begin{acknowledgments}
\noindent 
V. T. Z.  thanks Funda\c c\~ao de
Amparo \`a Pesquisa do Estado de S\~ao Paulo - FAPESP, Grant No. 
2011/18729-1,
Conselho Nacional de Desenvolvimento Cient\'\i fico e Tecnol\'ogico - CNPq, 
Brazil, Grant No. 308346/2015-7, and Coordena\c{c}\~ao de Aperfei\c{c}oamento 
de Pessoal de N\'\i vel Superior - CAPES, Brazil,  Grant
No.~88881.064999/2014-01.

\end{acknowledgments}

\appendix

\section{Equations of structure in dimensionless form}
\label{eq_adimen_politropico}

For the numerical calculations, the equations of structure are written in a
dimensionless form. In Ref.~\cite{ALZ} this was done for non-relativistic 
polytropes. Here we present the normalized equations of structure for
relativistic polytropes only. 

The normalized radial coordinate $\xi(r)$ is defined by 
\begin{equation}\label{radius_adimensional}
\xi=r\sqrt{4\pi\delta_{c}}, 
\end{equation}
where we have put $c=1$ and also $G=1$.
Similarly, the electric charge function 
$q(r)$, the mass function $m(r)$, and the rest-mass density function 
$\delta(r)$ are replaced by the new normalized variables, $\mu(\xi)$, $v(\xi)$ 
and $\vartheta(\xi)$, respectively, defined by
\begin{eqnarray}
& & \mu(\xi)=q(r)\frac{\sqrt{4\pi\delta_{c}}}{\xi^{2}} ,  \label{q_u} \\
& & v(\xi)=m(r)\sqrt{4\pi\delta_{c}} ,  \label{m_v} \\
& & \vartheta(\xi)=\left(\frac{\delta(r)}{\delta_{c}}\right)^{\gamma},
\label{rho_thet}
\end{eqnarray}
where $\delta_c$ represents the central rest mass density. 
In terms of $\vartheta$ and $\delta_c$, the
pressure and the energy density are now given by relations $%
p(r)=\omega\delta_{c}^{\gamma}\vartheta(\xi)$ and $\rho(r)=\delta_c\left[%
\vartheta^{1/\gamma}+\omega\delta_{c}^{\gamma-1}\vartheta/(\gamma-1)%
\right] $, respectively.

With these new variables, the equations of structure
\eqref{continuidad da carga}, \eqref{continuidad de la 
masa}, and \eqref{tov} provide,
\begin{eqnarray}
\hspace*{-.6cm}\frac{d\mu}{d\xi}&=&-\frac{2\mu}{\xi}+\frac{
\alpha\vartheta^{1/\gamma}
+\alpha(\gamma-1)^{-1}\omega\delta_{c}^{\gamma-1}\vartheta}{\sqrt{1-\dfrac{2v
}{\xi} +\xi^2 \mu^2}}, \hskip .9cm\label{u1} \\
\hspace{-0.6cm} \frac{d v}{d\xi}&\!=&\!\left[\vartheta^{1/\gamma} +\dfrac{
\omega\delta_{c}^{\gamma-1}\vartheta}{(\gamma-1)}\right] \left[\xi^{2}+\dfrac{
\alpha\xi^3 \mu}{\sqrt{1-\dfrac{2v}{\xi}+\xi^2 \mu^2}}\right],\;\;
\label{upsilon1} \\
\frac{d\vartheta}{d\xi}&=&-\xi\left[
a\vartheta+\omega^ { -1 }
\delta_{c}^{1 -\gamma}\vartheta^{1/\gamma}\right]\left[\frac{
\omega\delta_{c}^{\gamma-1}\vartheta-\mu^{2}+\dfrac{v} {\xi^{3}}}{1-\dfrac{2v
}{\xi}+\xi^2 \mu^2}\right] \notag \;\;\;\;\\
&&\hskip .5cm+\frac{\alpha
\mu\omega^{-1}\delta_{c}^{1-\gamma}\vartheta^{1/\gamma}+\alpha(
\gamma-1)^{-1}u\vartheta}{ \sqrt{1-\dfrac{2v}{\xi}+\xi^2 \mu^2}}, 
\label{theta1}
\end{eqnarray}
where Eq.~\eqref{densicarga_densimasa} was used to eliminate the charge 
density.

In order to get an equilibrium solution, the coupled equations 
\eqref{u1}--\eqref{theta1} are solved simultaneously, through numerical 
integration. 
After determining  $\mu(\xi)$, $v(\xi)$, and $\vartheta(\xi)$, the other 
function $B(\xi)$ and $A(\xi)$  are found from the equations
\begin{eqnarray}
\frac{dB}{d\xi}&=&2\xi B\left[\frac{\omega\delta_{c}^{\gamma-1}\vartheta-\mu^2+
\dfrac{v} {\xi^3}}{1-\dfrac{2v}{\xi}+\xi^2\mu^2}\right], \label{eq_gtt_adi} \\
A^{-1}&=&1-\frac{2v}{\xi}+\xi^2 \mu^2.
\end{eqnarray}

The boundary conditions assumed at the center of the sphere 
($\xi=0$) are: $\mu(0)=0$, $v(0)=0$, and $
\vartheta(0)=1$. The value of $\xi$ at the surface of the object is
determined by the condition $\vartheta(\xi_s)=0$,
where $\xi_s$ is identified as
the normalized radius at the surface of the sphere. 
The integration of Eqs. \eqref{u1}, \eqref{upsilon1}, and 
\eqref{theta1} is stopped when the value of $\vartheta$ changes
sign, from positive to negative. Once obtained the corresponding
values of $\xi_s$, $\mu(\xi_s)$, and $v(\xi_s)$, the values of the radius $R$, 
the mass $M$, and the charge $Q$ of the object are
determined using relations \eqref{radius_adimensional},
\eqref{q_u}, and \eqref{m_v}, respectively.

\end{document}